\documentclass[twocolumn]{aastex61}
\usepackage{natbib}
\usepackage{outlines}
\usepackage{enumitem}
\setenumerate[1]{label=\arabic*.}
\setenumerate[2]{label=\Alph*)}
\setenumerate[3]{label=\roman*)}
\setenumerate[4]{label=\alph*)}
\setenumerate[5]{label=\arabic*)}

\received{Oct. 4, 2017}
\revised{Dec. 5, 2017}
\accepted{Dec. 6, 2017}

\shorttitle{Transit Light Source Effect}
\shortauthors{Rackham et al.}

\begin{document}

\title{The Transit Light Source 
\replaced{Problem: False planetary spectral features in high-precision transit spectroscopy and incorrect planetary densities due to stellar heterogeneity in M dwarfs}
{Effect: False spectral features and incorrect densities for M-dwarf transiting planets}
}

\correspondingauthor{Benjamin V. Rackham}
\email{brackham@as.arizona.edu}

\author{Benjamin V. Rackham}
\affiliation{Department of Astronomy/Steward Observatory, The University of Arizona, 933 N. Cherry Avenue, Tucson, AZ 85721, USA}
\affiliation{National Science Foundation Graduate Research Fellow.}
\affiliation{Earths in Other Solar Systems Team, NASA Nexus for Exoplanet System Science.}

\author{D\'aniel Apai}
\affiliation{Department of Astronomy/Steward Observatory, The University of Arizona, 933 N. Cherry Avenue, Tucson, AZ 85721, USA}
\affiliation{Department of Planetary Sciences, The University of Arizona, 1629 E. University Blvd, Tucson, AZ 85721, USA}
\affiliation{Earths in Other Solar Systems Team, NASA Nexus for Exoplanet System Science.}

\author{Mark S. Giampapa}
\affiliation{National Solar Observatory, 950 N. Cherry Avenue, Tucson, AZ 85719, USA}



\begin{abstract}

Transmission spectra are differential measurements that utilize stellar illumination to probe \deleted{the structure and composition of }transiting exoplanet atmospheres. Any spectral difference between the illuminating light source and the disk-integrated stellar spectrum due to \deleted{the presence of }starspots and faculae will be imprinted in the observed transmission spectrum. However, \deleted{despite their clear importance, }few constraints exist for the extent of photospheric heterogeneities in M dwarfs. Here, we \replaced{present a forward model to assess the range of}{model} spot and faculae covering fractions consistent with observed photometric variabilities for M dwarfs and \replaced{their}{the} associated \added{0.3--5.5~$\micron$} stellar contamination spectra. We find that large ranges of spot and faculae covering fractions are consistent with \replaced{observed variability levels. We also show that}{observations and} corrections \replaced{that assume}{assuming} a linear \replaced{correlation}{relation} between variability amplitude and covering fractions generally underestimate the \deleted{level of }stellar contamination.\deleted{present.} Using realistic estimates for spot and faculae covering fractions, we find \deleted{the }stellar contamination \deleted{signal }can be more than $10 \times$ larger than transit depth changes expected for atmospheric features in rocky exoplanets. \replaced{Also, integrating across photometric bands, we}{We also} find that stellar spectral contamination can lead to \deleted{significant }systematic errors in radius and therefore the derived density of small planets. In the \deleted{specific }case of the \object{TRAPPIST-1} system, we show that \deleted{instead of the assumed 1--2~\% spot coverage, }TRAPPIST-1's rotational variability is \deleted{in fact }consistent with spot covering fractions 
\replaced{$f_{spot} = 12^{+10}_{-11}\%$}
{$f_{spot} = 8^{+18}_{-7}\%$}
and faculae covering fractions 
\replaced{$f_{fac} = 65^{+7}_{-55}\%$}
{$f_{fac} = 54^{+16}_{-46}\%$}. 
The associated stellar contamination signals alter transit depths of the TRAPPIST-1 planets at wavelengths of interest for planetary atmospheric species by roughly 1--15 $\times$ the strength of planetary features, significantly complicating \replaced{James Webb Space Telescope}{\textit{JWST}} follow-up observations of this system. Similarly, we find that stellar contamination can lead to underestimates of bulk densities of the TRAPPIST-1 planets of 
\replaced{$12^{+11}_{-12} \%$}
{$\Delta(\rho) = -3^{+3}_{-8} \%$ }, 
thus leading to overestimates of their volatile contents.

\end{abstract}




\section{Introduction} \label{sec:intro}

Transmission spectroscopy, the multi-wavelength study of transits that reveals the apparent size of the exoplanet as a function of wavelength \citep[e.g.,][]{Seager2000, Brown2001}, provides the best opportunity to study the atmospheres of small and cool exoplanets in the coming decades. During a transit, exoplanets appear larger at some wavelengths due to absorption or scattering of starlight by their atmospheres. The scale of the signal depends inversely on the square of the stellar radius \citep{Miller-Ricci2009}, prompting a focus on studying exoplanets around M dwarf stars.

A rapidly growing number of exciting M dwarf exoplanet systems hosting super-Earth and Earth-mass planets have been discovered to date, including \object{GJ 1132b} \citep{Berta-Thompson2015}, \object{LHS 1140b} \citep{Dittmann2017}, and the \object{TRAPPIST-1} system \citep{Gillon2016, Gillon2017,Luger2017}, an ultracool dwarf only 12 parsecs away hosting a system of seven transiting Earth-sized planets. The low densities of the TRAPPIST-1 planets may indicate high volatile contents, and as many as three of them may have surface temperatures temperate enough for \added{long-lived} liquid water to exist \citep{Gillon2017}. \added{Frequent flaring \citep{Vida2017} and strong XUV radiation from the host star \citep{Wheatley2017}, however, can lead to significant water loss for these planets \citep{Bolmont2017}, and 3D climate modeling suggests TRAPPIST-1e provides the best opportunity for present-day surface water and an Earth-like temperature in the system \citep{Wolf2017}.}

\added{While M dwarf exoplanets provide an excellent opportunity to study small and cool exoplanets \citep{Barstow2016}, they also represent a significant challenge. Spots with covering fractions as low as 1\% on M dwarfs introduce radial velocity jitter that can mask the presence of habitable zone Earth-sized exoplanets \citep{Andersen2015}. Variability monitoring suggests 1--3\% of M dwarfs have spot covering fractions of 10\% or more \citep{Goulding2012}. In addition to radial velocity jitter, unocculted spots also introduce errors in wavelength-dependent planetary radii recovered from transit observations \citep[e.g.,][]{Pont2008}.} \replaced{In studying transiting exoplanets, it is crucial to delineate the magnitude and sources of error in radii estimates, as}{Given the dependence of density calculations on measurements of exoplanet radii ($\rho \propto R^{-3}$), any} errors in radius determinations are amplified by a factor of 3 in the estimate of the exoplanet bulk density\replaced{. Combined with the measurement uncertainty in exoplanet mass, errors in exoplanet radii}{ and} can lead to significant consequences for the development of accurate exoplanet models.

\begin{figure*}[!htbp]
\includegraphics[width=\linewidth]{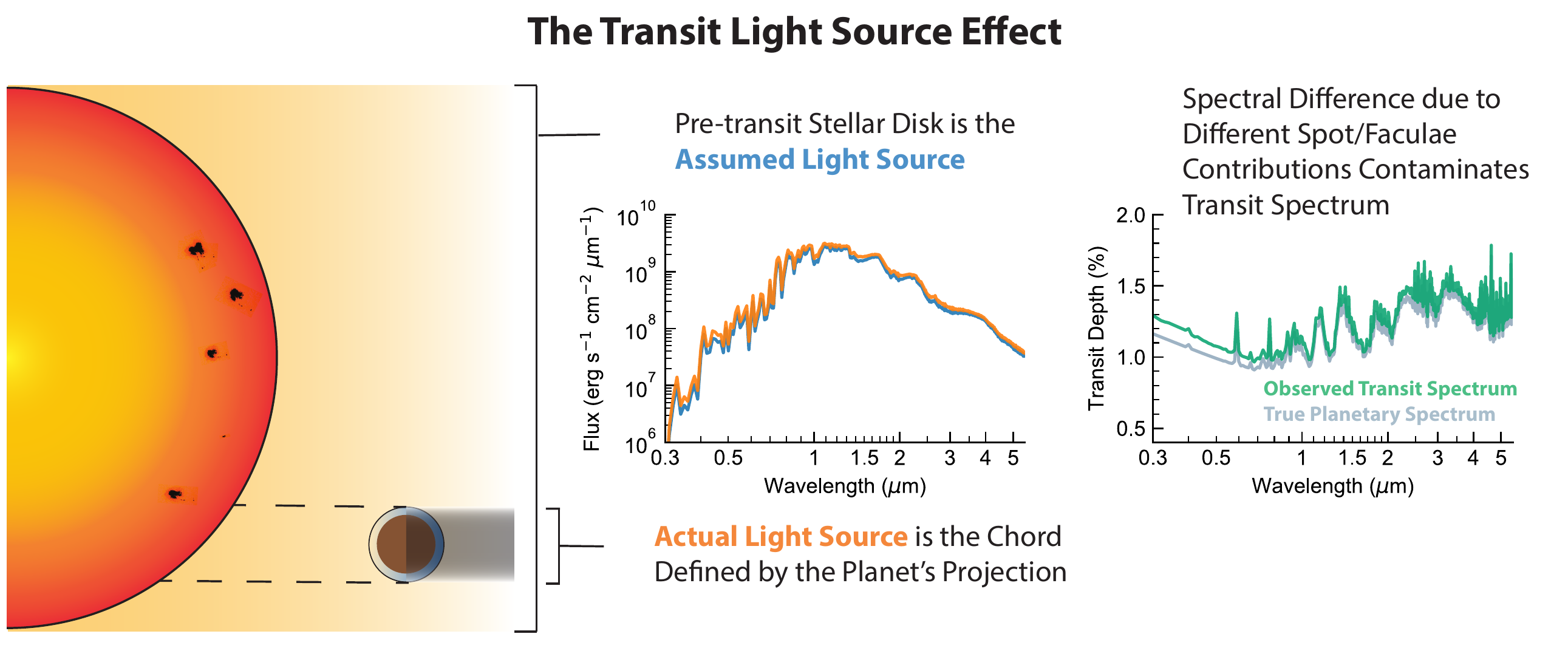}
\caption{A Schematic of the Transit Light Source \replaced{Problem}{Effect}. During a transit, exoplanet atmospheres are illuminated by the portion of a stellar photosphere immediately behind the exoplanet from the point of view of the observed. Changes in transit depth must be measured relative to the spectrum of this light source. However, the light source is generally assumed to be the disk-integrated spectrum of the star. Any differences between the assumed and actual light sources will lead to apparent variations in transit depth. \label{fig:TLSP}}
\end{figure*}

\replaced{Chief among the sources of radii errors is what}{Unocculted spots are one manifestation of a generic issue with transit observations that} we term the \textit{transit light source \replaced{problem}{effect}} (Figure~\ref{fig:TLSP}): Any transmission spectroscopic measurement relies on measuring the difference between the {\em incident} and {\em transmitted} light to identify the absorbers present in the media studied \added{\citep[e.g.,][]{Seager2000}}. The level of accuracy with which the incident spectrum is known will directly determine the level of accuracy with which the transmitted light is understood. In the transiting exoplanet case, the incident light is measured by observing the disk-integrated stellar spectrum before the transit \added{\citep[e.g.,][]{Brown2001}}, the assumption being that the disk-integrated spectrum is identical to the light incident on the planetary atmosphere\deleted{ in transmission spectroscopy studies}. However, this is only an approximation: the planet is not occulting the entire stellar disk but only a small \replaced{chord}{region within the transit chord at a given time}{. Thus, the light source for the transmission measurement is a small time-varying annulus within the stellar disk defined by the planet's projection}, the spectrum of which may differ significantly from the disk-averaged spectrum. Such differences are expected due to the fact that stellar atmospheres are rarely perfectly homogeneous, \replaced{i.e., in almost all cases the spectrum of any arbitrary small part of the stellar disk will {\em not} be identical to the spectrum of the entire disk.}{as illustrated by spatially resolved observations of the Sun \citep[e.g.,][]{Llama2015, Llama2016}.} Cool stellar spots (umbra and penumbra), hot faculae, and even latitudinal temperature gradients will result in a spectral mismatch, even if some of these will not be evident in broad-band photometric light curves. \deleted{For example, an axisymmetric latitudinal temperature gradient (e.g., cooler polar than equatorial temperatures) will result in no photometric variations but very significant spectral differences within the stellar disk. In essence, an exoplanet transit spatially samples the stellar disk along a particular transit chord that, for low obliquities, crosses at a specific latitude. In the case of oblique exoplanet orbits, the cumulative transmission spectrum will include contributions from the inhomogeneous stellar disk spanning a range of latitudes.}

The Sun displays a clear latitudinal dependence of active regions that gives rise to the so-called butterfly diagram \citep{Maunder1922, Babcock1961, Mandal2017}. Transiting exoplanets have been proposed to be tools to probe the latitudinal and temporal distributions of active regions in other stars \citep{Dittmann2009, Llama2012}. \added{High-resolution transit observations can be used to spatially resolve the emergent stellar spectrum along the transit path of the planet \citep{Cauley2017, Dravins2017a, Dravins2017b}.} \replaced{Regions of concentrated magnetic flux emerge preferentially at low latitudes where transit chords for low obliquities predominately occur. At high exoplanet obliquities a range of latitudes is sampled with activity encountered that may be substantially modulated by the underlying rotation of the active host star.}{\citet{Morris2017} recently utilized the highly misaligned exoplanet \object{HAT-P-11b} to probe the starspot radii and latitudinal distribution of its K4 dwarf host star and found that, much like the Sun, spots on \object{HAT-P-11} emerge preferentially at two low latitudes. In general, however, orbital planes of transiting exoplanets tend to be more aligned with stellar rotation axes than that of HAT-P-11b, displaying obliquities of $\lesssim 20 \degr$ \citep{Winn2017}. To complicate matters further, unlike the Sun and HAT-P-11, M dwarfs may exhibit spots at all latitudes \citep{Barnes2001a}. Thus, stellar latitudes sampled by transit chords may not provide a representative picture of photospheric active regions.} 

Correcting transmission spectra for photospheric heterogeneities within the transit chord, such as spots \citep[e.g.,][]{Pont2013, Llama2015} and faculae \citep{Oshagh2014}, is \replaced{relatively straightforward}{possible}, provided they are large enough to produce an observable change in the light curve during the transit. Modulations in the shape of the transit light curve can be used to constrain the temperature \citep{Sing2011} and size \citep{Beky2014} of the occulted photospheric feature, which determine its contribution to the transmission spectrum, or more simply, time points including the crossing event may be excluded from the transit fit \citep[e.g.,][]{Pont2008, Carter2011, Narita2013}.

Unocculted heterogeneities, however, represent a more pathological manifestation of the transit light source \replaced{problem}{effect} because they do not produce temporal changes in the observed light curve. Previous attempts to correct for unocculted photospheric features have largely relied on photometric monitoring of the exoplanet host star to ascertain the extent of photospheric heterogeneities present \citep{Pont2008, Pont2013, Berta2011, Desert2011, Sing2011, Knutson2012, Narita2013, Nascimbeni2015, Zellem2015}. This approach is limited in two respects: 1) rotational variability monitoring traces only the non-axisymmetric component of the stellar heterogeneity \added{\citep{Jackson2012}}, i.e., any persistent, underlying level of heterogeneity will not be detectable with variability monitoring; and 2) the source of the variability is commonly assumed to be a single giant spot, the size of which scales linearly with the variability amplitude\added{, an assumption which provides only a lower limit on the extent of active regions}.

\citet{Zellem2017} present a novel method to remove relative changes in the stellar contribution to individual transits utilizing the out-of-transit data flanking each transit observation. The strength of this approach lies in \deleted{fact }that it does not require additional measurements to provide a relative correction for differences in spot and faculae covering fractions between observations. However, as with other variability-based techniques, this approach cannot correct for any persistent level of spots or faculae that may be present in all observations and can strongly alter transmission spectra \citep{McCullough2014, Rackham2017}.

Useful constraints on spot and faculae covering fractions are hindered by observational and theoretical limits on our knowledge of stellar photospheres. On the Sun, the disk passage of sunspots can produce relative declines in the solar total irradiance in the range of $\sim$~0.1\% to 0.3\% \citep[e.g.,][]{Kopp2005}. By contrast, \replaced{typical M dwarfs}{field mid-to-late M dwarfs ($M < 0.35~M_{\sun}$) with detectable rotation periods} display rotational modulations with semi-amplitudes of 0.5--1.0\% \citep{Newton2016}, corresponding to peak-to-trough variability full-amplitudes of 1--2\%. Thus, variability amplitudes in M dwarfs are roughly an order of magnitude larger than in the Sun. 

Despite the clear importance of constraining spot and faculae covering fractions for exoplanet host stars, a systematic attempt to connect observed variabilities to covering fractions and thus stellar contamination signals is absent in the literature on transmission spectroscopy.

In this work, we employ a forward modeling approach to explore the range of spot covering fractions consistent with observed photometric variabilities for \added{field} M dwarf stars and their associated effects on \added{visual and near-infrared (0.3--5.5~$\micron$)} planetary transmission spectra. In Section~\ref{sec:methods}, we detail our model for placing constraints on spot and faculae covering fractions and their associated stellar contamination spectra. Section~\ref{sec:results} provides the modeling results. We place our results in the context of observational attempts to constrain stellar heterogeneity and examine their impact on transmission spectra and density estimates of M dwarf exoplanets in Section~\ref{sec:discussion}, including a focused discussion of the TRAPPIST-1 system. Finally, we summarize our conclusions in Section~\ref{sec:conclusions}.

\section{Methods} \label{sec:methods}

\subsection{Synthetic stellar spectra} \label{sec:component_spectra}

We employed the PHOENIX \citep{Husser2013} and DRIFT-PHOENIX \citep{Witte2011} stellar spectral model grids to generate spectra for the \added{immaculate} photospheres, spots, and faculae of main sequence M dwarfs with spectral types from M0V to M9V. 
\added{Both model grids are based on the stellar atmosphere code PHOENIX \citep{Hauschildt1999}, with the DRIFT-PHOENIX model grids including additional physics describing the formation and condensation of mineral dust clouds \citep{Woitke2003, Woitke2004, Helling2006, Helling2008, Helling2008a, Witte2009} that is applicable to late M dwarfs and brown dwarfs.}
We considered models with solar metallicity ([Fe/H] = 0.0) and no alpha-element enrichment ([$\alpha$/Fe] = 0.0). We linearly interpolated between spectra in the grids to produce \added{0.3--5.5~$\micron$} model spectra for the surface gravities and temperatures we required.

\replaced{For simplicity, we neglect the dependence of faculae spectra on magnetic field strength and limb distance \citep{Norris2017} and save those complications for consideration in future work.}{The implicit assumption with this approach is that the emergent spectrum from distinct components of a stellar photosphere, such as the immaculate photosphere, spots, and faculae, can be approximated by models of disk-integrated stellar spectra of different temperatures. This approximation is commonly used in transit spectroscopy studies to constrain the contribution of unocculted photospheric heterogeneities to exoplanet transmission spectra \citep{Pont2008, Sing2011, Huitson2013, Jordan2013, Pont2013, Fraine2014, Sing2016, Rackham2017}. However, this simplification neglects the dependence of the spectra of photospheric heterogeneities on magnetic field strength and limb distance, both of which modulate the emergent spectra of magnetic surface features \citep{Norris2017}. Nonetheless, we adopt the simplifying assumption of parameterizing component spectra by temperature and note that future efforts may benefit from the increased realism of 3D magnetohydrodynamics models.}

\begin{deluxetable}{ccccc}[!tbp]
\tabletypesize{\small}
\tablecaption{Adopted stellar parameters \label{tab:stars}}
\tablehead{
		   \colhead{Sp. Type}                 &
		   \colhead{$T_{phot}$ (K)}           &
		   \colhead{$T_{spot}$ (K)}           &
           \colhead{$T_{fac}$ (K)}            &
           \colhead{log g (cgs)}                    
		  }
\startdata
M0V & 3800 & 3268 & 3900 & 4.7 \\
M1V & 3600 & 3096 & 3700 & 4.7 \\
M2V & 3400 & 2924 & 3500 & 4.8 \\
M3V & 3250 & 2795 & 3350 & 4.9 \\
M4V & 3100 & 2666 & 3200 & 5.3 \\
M5V & 2800 & 2408 & 2900 & 5.4 \\
M6V & 2600 & 2236 & 2700 & 5.6 \\
M7V & 2500 & 2150 & 2600 & 5.6 \\
M8V & 2400 & 2064 & 2500 & 5.7 \\
M9V & 2300 & 1978 & 2400 & 5.6 \\
\enddata
\tablecomments{\added{The photosphere temperature $T_{phot}$, spot temperature $T_{spot}$, facula temperature $T_{fac}$, and surface gravity $log~g$ we adopt for each M dwarf spectral type are listed.}}
\end{deluxetable}

Table~\ref{tab:stars} lists our adopted stellar parameters. For each spectral type, we calculated the surface gravity $g$ from the stellar masses and radii summarized by \citet{Kaltenegger2009} and adopted the stellar effective temperature from that same work as the photosphere temperature $T_{phot}$. Following \citet{Afram2015}, we adopted the relation $T_{spot} = 0.86 \times T_{phot}$, in which $T_{spot}$ is the spot temperature. We adopted the scaling relation $T_{fac} = T_{phot} + 100$~K \citep{Gondoin2008} for the facula temperature. Although there are uncertainties in the scaling relations of starspots and faculae, we do not expect our general results to be sensitive to the adopted relations. The temperature ranges of the spectral grids allowed us to simulate photosphere, spot, and facula spectra for spectral types M0V--M5V with the PHOENIX model grid and M5V--M9V with the DRIFT-PHOENIX model grid. 

\subsection{Spot covering fraction and variability amplitude relation \label{sec:methods_fspot}}

We explored the range of spot covering fractions consistent with an observed 1\% I-band variability full-amplitude for each spectral type. We modeled the stellar photosphere using a rectangular grid with a resolution of $180 \times 360$ pixels.  We initialized the model with an immaculate photosphere, setting the value of each resolution element to the flux of the photosphere spectrum integrated over Bessel I-band response \footnote{\url{http://www.aip.de/en/research/facilities/stella/instruments/data/johnson-ubvri-filter-curves}}. Likewise, when adding spots or faculae to the model, we utilized the integrated I-band fluxes of their respective spectra.

We considered four cases of stellar heterogeneities by varying two parameters:  spot size and the presence or absence of faculae. In terms of spot size, we examined cases with smaller and larger spots, which we deem the ``solar-like spots" and ``giant spots" cases. In the solar-like spots case, each spot 
\replaced{covered an approximately circular $2 \degr$ area}
{had a radius of $R_{spot} = 2 \degr$}, covering 400~ppm of the projected hemisphere and representing a large spot group on the Sun \citep{Mandal2017}. In the giant spots case, each spot 
\replaced{covered a $7 \degr$ area}
{had a radius of $R_{spot} = 7 \degr$}, covering 5,000~ppm of the projected hemisphere and corresponding roughly to the largest spots detectable on active M~dwarfs through molecular spectropolarimetry \citep{Berdyugina2011}. For cases with faculae, we included faculae at a facula-to-spot area ratio of 10:1, following observations of the active Sun \citep{Shapiro2014}. Thus, the stellar heterogeneity cases we considered were the following:  solar-like spots, giant spots, solar-like spots with faculae, and giant spots with faculae.

\begin{figure*}[!htbp]
\includegraphics[width=\linewidth]{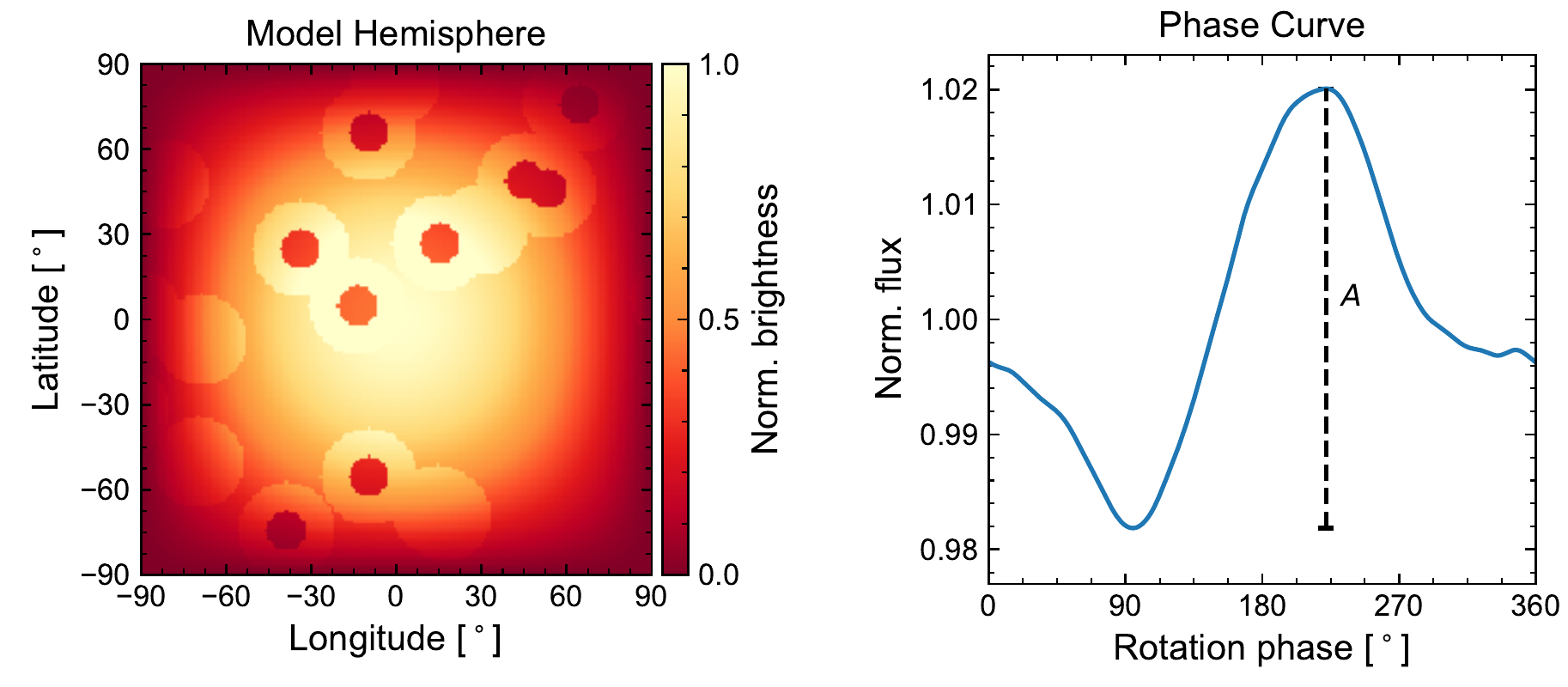}
\caption{\added{Example of a} model stellar photosphere and variability amplitude determination. The left panel shows one hemisphere of \replaced{a}{an example} model photosphere with \deleted{3\% coverage of} giant spots \added{and facular regions} after applying a double cosine weighting kernel. The right panel displays the phase curve produced by summing the hemispheric flux over one complete rotation of the model. The vertical dashed line illustrates the variability full-amplitude $A$, \deleted{which is} defined as the difference between the maximum and minimum normalized flux \added{, which is $\sim 4$\% in this case}. \label{fig:model}}
\end{figure*}

For each spectral type and stellar heterogeneity case, we examined the dependence of the variability on the spot covering fraction through an iterative process. In each iteration, we added a spot to the model photosphere at a randomly selected set of coordinates\footnote{We assumed no latitudinal dependence for the spot distribution. This assumption is good for active M dwarf stars \citep{Barnes2001a, Barnes2001b}, but may not hold for earlier spectral types \citep{Morris2017}. We will examine the additional complication of latitudinal dependence of photospheric features in a future paper.}, recorded the spot covering fraction, and generated a phase curve.  In cases including faculae, we added half of the facular area at positions adjacent to the spot and half in a roughly circular area at another randomly selected set of coordinates. We allowed spots to overwrite faculae but not vice versa in successive iterations to ensure the spot covering fraction increased monotonically. We generated a phase curve by applying a double cosine weighting kernel to one hemisphere of the rectangular grid ($180 \times 180$ pixels), summing the flux, and repeating the process for all 360 x-coordinates (``latitudes'') in the model (Figure~\ref{fig:model}). We recorded the variability full-amplitude $A$ as the difference between the minimum and maximum normalized flux in the phase curve at each iteration. \added{This approach assumes the stellar rotation axis is aligned well with the plane of the sky. As the presence of transits ensures that the planetary orbital plane is nearly edge-on, and obliquities between the stellar rotation axis and planetary orbital plane are generally $\lesssim 20 \degr$ \citep{Winn2017}, this assumption is good for most transiting exoplanet systems.} Following this procedure, we added spots to the immaculate photosphere iteratively until reaching 50\% spot coverage.

We repeated this procedure 100 times for a given set of stellar parameters and heterogeneity case to examine the central tendency and dispersion in modeling results. In each trial, we recorded the minimum spot covering fraction that produced a variability full-amplitude of 1\% ($A=0.01$). Using the results of the 100 trials, we calculated the mean spot covering fraction $f_{spot ,mean}$ corresponding to $A=0.01$ and its standard deviation. We defined the spot covering fractions $1 \sigma$ below and above the mean as $f_{spot, min}$ and $f_{spot, max}$, respectively.

As spots were allowed to overwrite faculae in our model but not vice versa, the facula-to-spot area ratio drifted from its original 10:1 value as spots were added to the model. Thus a distribution of facular-covering fractions existed for each spot covering fraction of interest. Accordingly, to quantify the central tendency and dispersion in results for models including faculae, we calculated the mean and standard deviation of facular-covering fractions in the 100 trials for each spot covering fraction of interest.  We defined $f_{fac, mean}$ as the mean facular-covering fraction corresponding to $f_{spot, mean}$, $f_{fac, min}$ as the mean facular-covering fraction corresponding to $f_{spot, min}$ minus one standard deviation of that distribution, and $f_{fac, max}$ as the mean facular-covering fraction corresponding to $f_{spot, max}$ plus one standard deviation of that distribution.

\subsection{Model for stellar contamination spectra} \label{sec:contamination_spectra_model}

Using the spot and faculae covering fractions determined through our variability modeling, we modeled the effect of stellar heterogeneity on observations of \added{visual and near-infrared (0.3--5.5~$\micron$)} exoplanet transmission spectra. We utilized the Composite Photosphere and Atmospheric Transmission (CPAT) model described in \citet{Rackham2017} and the spectra described in Section~\ref{sec:component_spectra}. Recasting Equation~(11) of \citet{Rackham2017} in terms of transit depths and simplifying terms, we find
\begin{equation}
D_{\lambda, obs}= \frac{D_{\lambda}}
                       {1 - f_{het}(1 - \frac{F_{\lambda, het}}{F_{\lambda, phot}})},
\label{eq:CPAT}
\end{equation}
in which $D_{\lambda, obs}$ is the observed transit depth, $D_{\lambda}$ is the nominal transit depth (i.e., the square of the true wavelength-dependent planet-to-star radius ratio), $F_{\lambda, phot}$ is the spectrum of the photosphere, $F_{\lambda, het}$ is the spectrum of a photospheric heterogeneity (i.e., spots or faculae), and $f_{het}$ is the fraction of the projected stellar disk covered by the heterogeneity \citep[see also][Equation~(1)]{McCullough2014}. This formalism assumes the transit chord can be described well by $F_{\lambda, phot}$, which is the case for transits of an immaculate photosphere.
\replaced{or those of sufficient quality that spot- and faculae-crossings can be detected in the light curve and removed from the analysis.}
{It also applies to transit observations in which the amplitude of the spot- or faculae-crossing event is larger than the observational uncertainty, thus enabling the parameters of the photospheric heterogeneity to be modeled \citep[e.g.,][]{Sanchis-Ojeda2011, Huitson2013, Pont2013, Tregloan-Reed2013, Scandariato2017} or the affected portion of the light curve to be removed from the analysis \citep[e.g.,][]{Pont2008, Carter2011, Narita2013}.}

The denominator on the right side of Equation~(\ref{eq:CPAT}) represents the signal imprinted on the observed transit depth by the stellar heterogeneity.  It is a multiplicative change to the transit depth independent of the exoplanet transmission spectrum. Therefore, by dividing Equation~(\ref{eq:CPAT}) by $D_{\lambda}$, we can define the term
\begin{equation}
\epsilon_{\lambda, het}= \frac{1}
                         {1 - f_{het}(1 - \frac{F_{\lambda, het}}{F_{\lambda, phot}})},
\label{eq:CS}
\end{equation}
which we refer to as the contamination spectrum hereafter. Approaching the problem in this way allows contamination spectra to be calculated for different stellar parameters and applied to any exoplanetary transmission spectrum of interest.

Equation~(\ref{eq:CS}) holds when the stellar disk can be described by two spectral components (i.e., immaculate photosphere and spots). In the case that spots and faculae are present in the stellar disk, the expression becomes
\begin{equation}
\epsilon_{\lambda, s+f}= \frac{1}
                         {1 - f_{spot}(1 - \frac{F_{\lambda, spot}}{F_{\lambda, phot}}) - f_{fac}(1 - \frac{F_{\lambda, fac}}{F_{\lambda, phot}})},
\label{eq:CS_sf}
\end{equation}
in which the $\epsilon_{\lambda, s+f}$ is the contamination spectrum produced by the combination of unocculted spots and facular regions and the subscripts ``spot'' and ``fac'' refer to spots and faculae, respectively. 

\section{Results} \label{sec:results}

\subsection{Spot covering fraction and variability amplitude relation}

\begin{deluxetable*}{ccccc}[p]
\tabletypesize{\small}
\tablecaption{Summary of variability modeling results by heterogeneity case \label{tab:results}}
\tablehead{
		   \colhead{Model Parameter}                &
           \multicolumn{4}{c}{Heterogeneity Case}  \\
           \cline{2-5}
           \colhead{}                              &
           \colhead{Giant Spots}                   &
		   \colhead{Solar-like Spots}              &
		   \colhead{Giant Spots + Faculae}         &
           \colhead{Solar-like Spots + Faculae}      
		  }
\startdata
Spot covering fraction, $f_{spot} (\%)$  & ${0.9}^{+1.3}_{-0.4}$ & ${12}^{+23}_{-6}$ & ${0.5}^{+0.8}_{-0.1}$ & ${14}^{+16}_{-7}$ \\
Faculae covering fraction, $f_{fac}(\%)$ & -                     & -               & ${4.6}^{+7.4}_{-1.4}$ & ${63}^{+5}_{-25}$ \\
Average transit depth change, $\overline{\epsilon}$ (\%) & 0.4   & 5.0   & -0.4     & 5.8 \\
Primary contributor to contamination spectrum            & Spots & Spots & Faculae & Spots \\
\enddata
\tablecomments{\added{Here we summarize key results from the four cases of stellar photosphere heterogeneities that we considered (see Section~\ref{sec:methods_fspot}). For each heterogeneity case, we provide the mean covering fractions of spots (and faculae, if possible) consistent with a 1\% I-band rotational variability across all spectral types considered. The error bars on these values refer to the means of the $f_{spot, min}$, $f_{spot, max}$, $f_{fac, min}$, and $f_{fac, max}$ parameters defined in Section~\ref{sec:methods_fspot}. Also included is the average transit depth change produced by all models across the 0.3--5.5~$\micron$ wavelength range and a qualitative assessment of the photospheric heterogeneity that dominates the contamination spectrum.}}
\end{deluxetable*}

\begin{figure*}[htbp!]
\includegraphics[width=\linewidth]{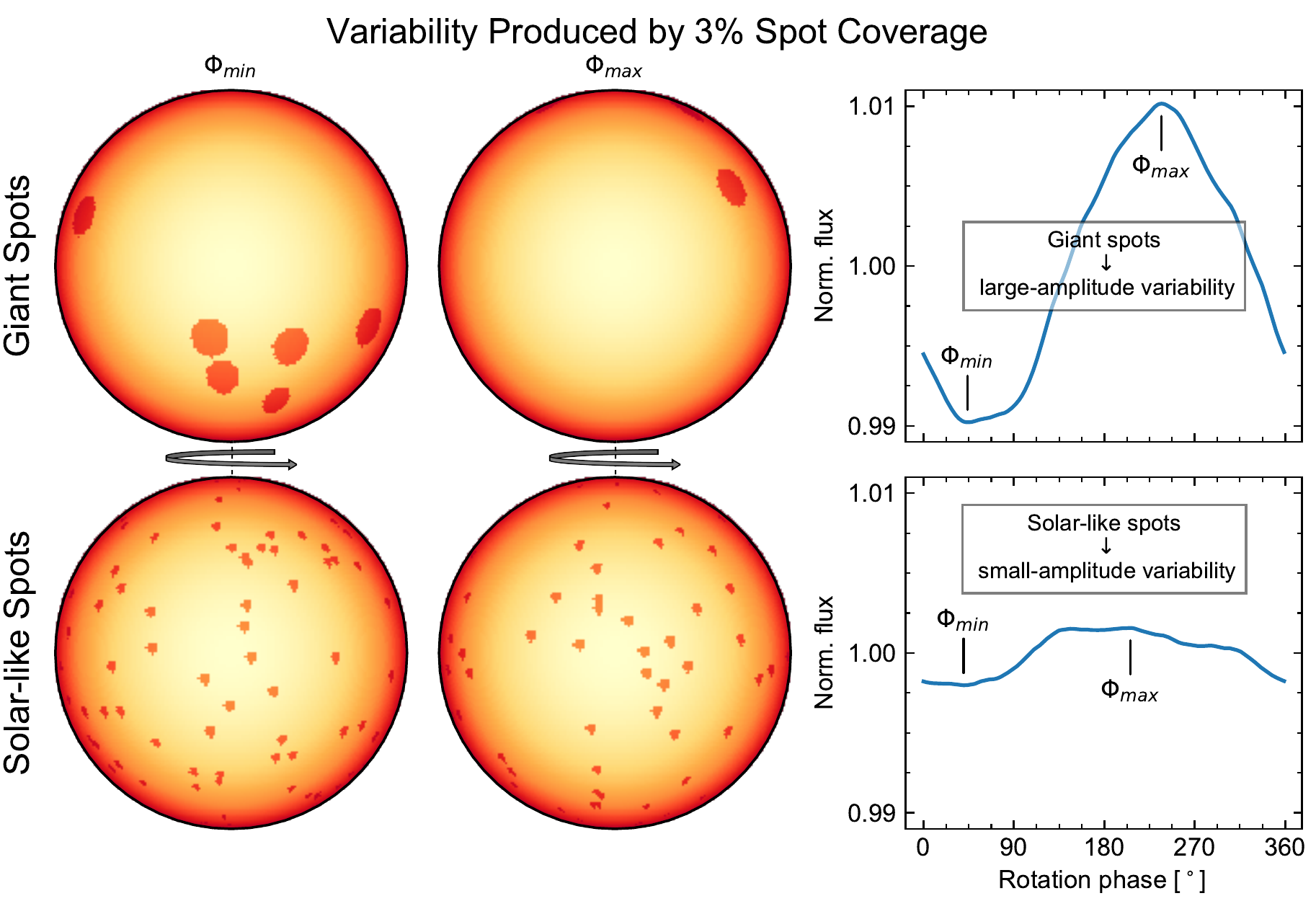}
\caption{Illustration of dependence of variability on spot size. The case of giant (solar-like) spots is shown on the top (bottom) row.  Both cases include 3\% spot coverage. The stellar hemispheres corresponding to the minima (maxima) of the phase curves are shown in the left (center) column. The right column shows the phase curves for the two cases, and the phases of illustrated hemispheres are indicated. \label{fig:illustration}}
\end{figure*}

\begin{deluxetable*}{cccccc}[htbp!]
\tabletypesize{\footnotesize}
\tablecaption{Filling factors and scaling coefficients determined by spots-only models \label{tab:filling_factors_spots_only}}
\tablehead{
		   \colhead{Sp. Type}             &
           \colhead{Model Grid}           &
		   \colhead{$f_{spot, min}$}      &
		   \colhead{$f_{spot, mean}$}     &
           \colhead{$f_{spot, max}$}      &
           \colhead{$C$\tablenotemark{a}}
		  }
\startdata
\multicolumn{6}{c}{Giant Spots} \\
\hline
M0V    & PHOENIX       & 0.007  & 0.011  & 0.026  & $0.084 \pm 0.030$ \\
M1V    & PHOENIX       & 0.006  & 0.011  & 0.029  & $0.092 \pm 0.036$ \\
M2V    & PHOENIX       & 0.006  & 0.011  & 0.028  & $0.104 \pm 0.041$ \\
M3V    & PHOENIX       & 0.006  & 0.010  & 0.022  & $0.100 \pm 0.038$ \\
M4V    & PHOENIX       & 0.005  & 0.008  & 0.019  & $0.105 \pm 0.043$ \\
M5V    & PHOENIX       & 0.006  & 0.010  & 0.023  & $0.092 \pm 0.035$ \\
M5V    & DRIFT-PHOENIX & 0.005  & 0.008  & 0.019  & $0.116 \pm 0.041$ \\
M6V    & DRIFT-PHOENIX & 0.005  & 0.008  & 0.015  & $0.106 \pm 0.040$ \\
M7V    & DRIFT-PHOENIX & 0.006  & 0.008  & 0.023  & $0.106 \pm 0.040$ \\
M8V    & DRIFT-PHOENIX & 0.006  & 0.009  & 0.023  & $0.106 \pm 0.041$ \\
M9V    & DRIFT-PHOENIX & 0.005  & 0.009  & 0.021  & $0.107 \pm 0.042$ \\
\hline
\multicolumn{6}{c}{Solar-like Spots} \\
\hline
M0V    & PHOENIX       & 0.10   & 0.20   & 0.50  &  $0.022 \pm 0.009$ \\
M1V    & PHOENIX       & 0.08   & 0.17   & 0.46  &  $0.024 \pm 0.009$ \\
M2V    & PHOENIX       & 0.07   & 0.12   & 0.42  &  $0.027 \pm 0.011$ \\
M3V    & PHOENIX       & 0.06   & 0.10   & 0.30  &  $0.029 \pm 0.010$ \\
M4V    & PHOENIX       & 0.05   & 0.11   & 0.30  &  $0.028 \pm 0.010$ \\
M5V    & PHOENIX       & 0.07   & 0.11   & 0.43  &  $0.027 \pm 0.010$ \\
M5V    & DRIFT-PHOENIX & 0.05   & 0.09   & 0.25  &  $0.032 \pm 0.012$ \\
M6V    & DRIFT-PHOENIX & 0.06   & 0.13   & 0.28  &  $0.029 \pm 0.010$ \\
M7V    & DRIFT-PHOENIX & 0.06   & 0.11   & 0.30  &  $0.030 \pm 0.011$ \\
M8V    & DRIFT-PHOENIX & 0.05   & 0.11   & 0.30  &  $0.028 \pm 0.011$ \\
M9V    & DRIFT-PHOENIX & 0.06   & 0.10   & 0.28  &  $0.028 \pm 0.010$ \\
\enddata
\tablenotetext{a}{\added{Scaling coefficient for square root scaling relation (Equation~\ref{eq:scaling_relation})}}
\tablecomments{\added{Section~\ref{sec:methods_fspot} provides the definitions of the spot covering fractions $f_{spot, min}$, $f_{spot, mean}$, and $f_{spot, max}$.}}
\end{deluxetable*}

We find that, for all M dwarf spectral types, an observed variability full-amplitude corresponds to a typically wide range of spot covering fractions. Averaging over all spectral types, Table~\ref{tab:results} provides a summary of the key results by heterogeneity case, including the spot and faculae covering fractions consistent with a 1\% variability full-amplitude, the average transit depth change over the full wavelength range studied, and the primary contributor (spots or faculae) to the stellar contamination spectrum.

For a given variability, we find that the spot covering fraction depends strongly on the spot size. Figure~\ref{fig:illustration} illustrates the origin of this dependence using two examples of model photospheres with $f_{spot}=3\%$. For a given spot covering fraction, the number density of spots is lower in the giant spots case than the solar-like spots case\replaced{. As the spots are randomly distributed throughout the photosphere, this increases the likelihood of developing a relative overdensity of spots.}{, leading to more concentrated surface heterogeneities and larger variability signals.} A single spot, for example, will always lead to rotational variability, while the variability signals from multiple spots positioned around the photosphere can add destructively, leading to a lower observed variability full-amplitude. The solar-like spots case demonstrates this effect. It includes the same spot covering fraction as the giant spots case but produces a markedly lower level of variability. While chromospheric diagnostics may be used to distinguish active and quiet stars, variability monitoring can only place a lower limit on the spot covering fraction. In effect, the rotational variability full-amplitude reflects only the non-axisymmetric component of the stellar heterogeneity. The axisymmetric component, which can be larger than the non-axisymmetric one, does not contribute to the observed variability but will affect transmission spectra.

\begin{figure*}[!htbp]
\includegraphics[width=\linewidth]{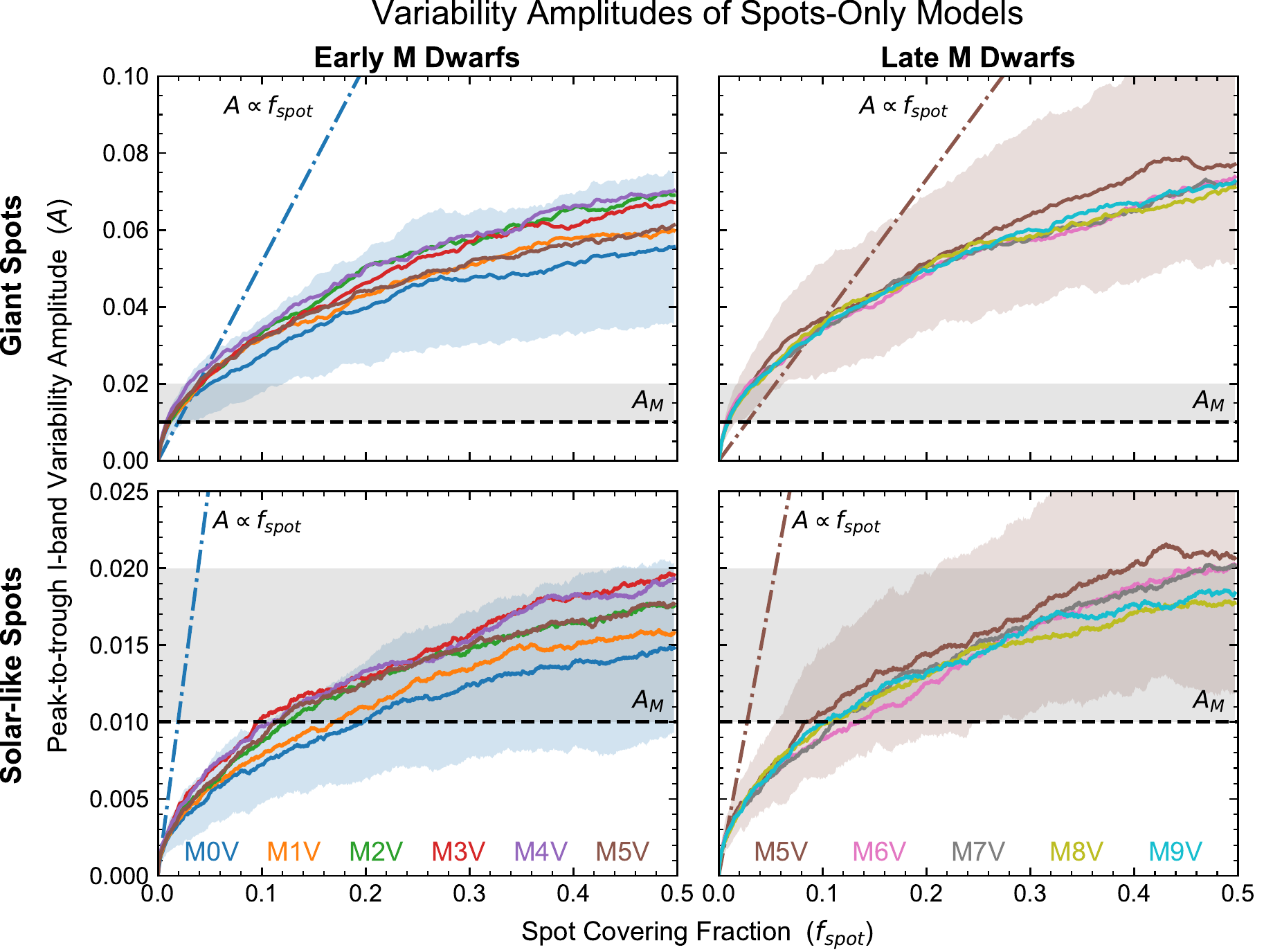}
\caption{Variability full-amplitudes as a function of spot covering fraction for the spots-only case. Results from PHOENIX and DRIFT-PHOENIX models are shown in the left and right columns, respectively. Results using giant and solar-like spots are shown in the top and bottom rows, respectively. Note the difference in vertical scale in the upper and lower panels. Solid lines give the mean spot covering fraction for each spectral type and the shaded region indicates the range encompassing 68\% of the model outcomes for the earliest spectral type in each plot, which is comparable to the dispersion in model outcomes for all other spectral types. Additionally, the expected relation if variability were a linear function of spot covering fraction is shown as a dash-dotted line, given the I-band photosphere and spot fluxes of the earliest spectral type. The horizontal dashed lines show the TRAPPIST-1 variability full amplitude and the gray shaded region highlights the range of typical M dwarf photometric modulations detected by \citet{Newton2016}. For each case, the variability grows asymptotically as a function of $f_{spot}$ and the linear relation generally underestimates $f_{spot}$. The dispersion in model outcomes demonstrates that a range of spot covering fractions corresponds to a given variability. \label{fig:fspot_variability}}
\end{figure*}

Figure~\ref{fig:fspot_variability} shows the results of our variability amplitude modeling efforts for the spots-only case. For each spectral type and spot size, the relationship between the spot covering fraction and observed \added{I-band} variability full-amplitude differs notably from a linear relation. Giant spots lead to overall larger variability full-amplitudes than solar-like spots do for the same spot covering fraction. With the exception of variability full-amplitudes $\le 1\%$ caused by giant spots, the linear relation is a poor approximation to the actual variability relation and underestimates the spot covering fraction. Instead, the variability full-amplitude grows asymptotically as a function of spot covering fraction. 

The shaded regions in Figure~\ref{fig:fspot_variability} illustrate the dispersion in the model outcomes. In effect, a given observed variability full-amplitude corresponds to a range of spot covering fractions, which widens further for larger variability full-amplitudes. Additionally, the solar-like spots case allows for still wider ranges of spot covering fractions. Table~\ref{tab:filling_factors_spots_only} provides the values of $f_{spot, min}$, $f_{spot, mean}$, and $f_{spot, max}$ (see Section~\ref{sec:methods_fspot}) \added{resulting from the set of 100 variability models we conducted} for each spectral type.

Following the apparent square root dependence of the variability full-amplitude $A$ on the spot covering fraction, we fit via least squares a scaling relation of the form
\begin{equation}
A = C \times f_{spot}^{0.5}
\label{eq:scaling_relation}
\end{equation}
to each set of models, in which $C$ is a scaling coefficient that depends on both the spot contrast and size. We find the relationship is approximated well by Equation~\ref{eq:scaling_relation} for all spots-only models. In Table~\ref{tab:filling_factors_spots_only}, we provide the fitted values of $C$ with an uncertainty determined by the 68\% dispersion in model outcomes (i.e., the shaded regions in Figure~\ref{fig:fspot_variability}).

\begin{figure*}[!htbp]
\includegraphics[width=\linewidth]{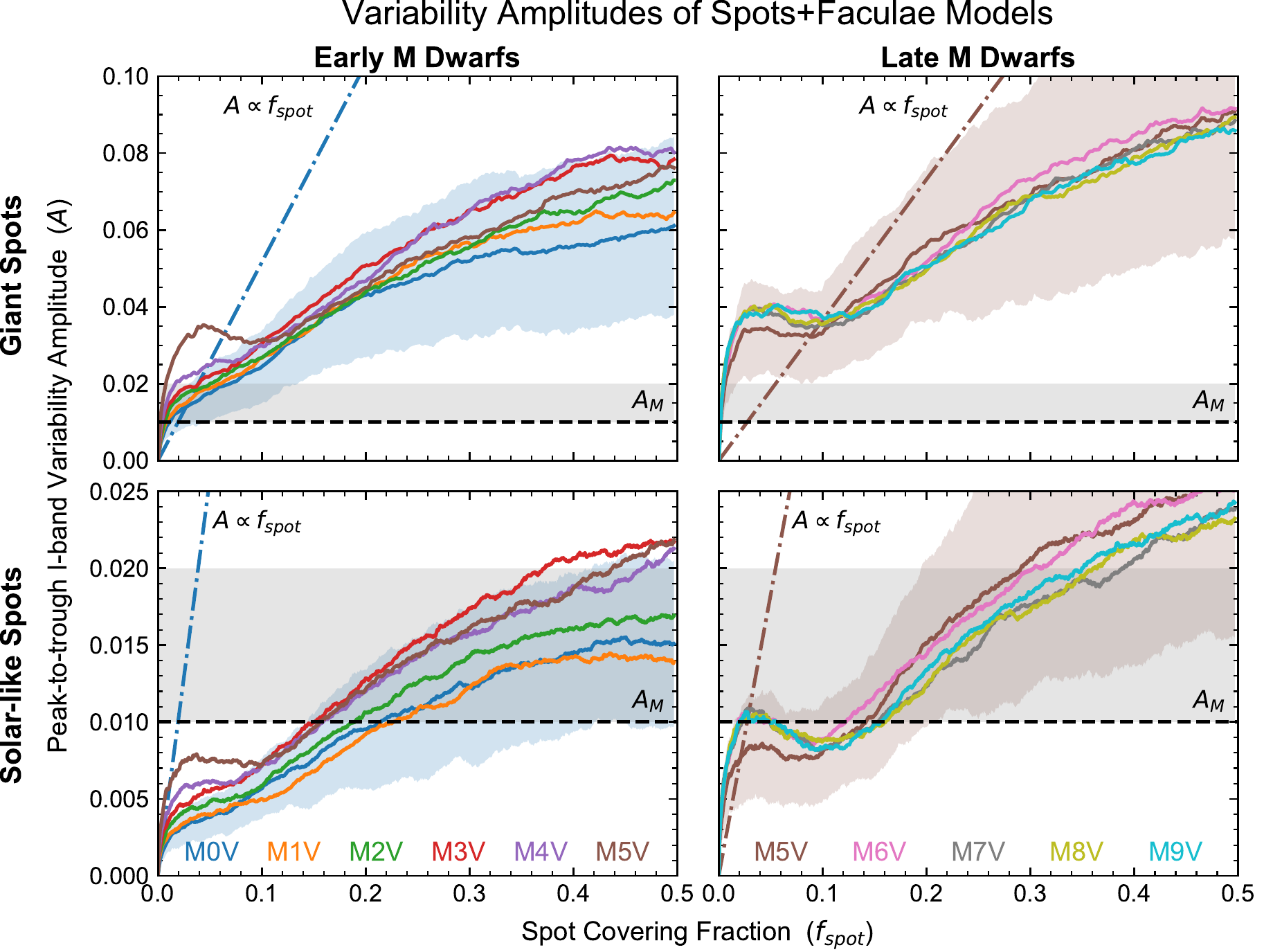}
\caption{Variability full-amplitudes as a function of spot covering fraction for the spot-and-faculae case. Note the difference in vertical scale in the upper and lower panels. Figure elements are the same as those in Fig.~\ref{fig:fspot_variability} \added{In contrast to the spots-only case, the addition of faculae in the variability modeling leads to larger variability amplitudes and a relative plateau in the relation for spot covering fractions $f_{spot} < 0.1$}. \label{fig:fspot_variability_faculae}}
\end{figure*}

\begin{deluxetable*}{cccccccc}[t]
\tabletypesize{\footnotesize}
\tablecaption{Filling factors determined by spots and faculae models \label{tab:filling_factors_spots_faculae}}
\tablehead{
		   \colhead{Sp. Type}             &
           \colhead{Model Grid}           &
		   \colhead{$f_{spot, min}$}      &
		   \colhead{$f_{spot, mean}$}     &
           \colhead{$f_{spot, max}$}      &
		   \colhead{$f_{fac, min}$}       &
		   \colhead{$f_{fac, mean}$}      &
           \colhead{$f_{fac, max}$}      
           }
\startdata
\multicolumn{8}{c}{Giant Spots and Faculae} \\
\hline
M0V    & PHOENIX       & 0.006  & 0.011  & 0.044  & 0.058  & 0.106  & 0.367  \\
M1V    & PHOENIX       & 0.005  & 0.008  & 0.029  & 0.047  & 0.082  & 0.265  \\
M2V    & PHOENIX       & 0.004  & 0.007  & 0.019  & 0.039  & 0.066  & 0.183  \\
M3V    & PHOENIX       & 0.003  & 0.005  & 0.016  & 0.032  & 0.049  & 0.158  \\
M4V    & PHOENIX       & 0.003  & 0.004  & 0.009  & 0.029  & 0.042  & 0.096  \\
M5V    & PHOENIX       & 0.003  & 0.003  & 0.005  & 0.024  & 0.030  & 0.051  \\
M5V    & DRIFT-PHOENIX & 0.003  & 0.003  & 0.004  & 0.023  & 0.029  & 0.046  \\
M6V    & DRIFT-PHOENIX & 0.003  & 0.003  & 0.004  & 0.023  & 0.025  & 0.040  \\
M7V    & DRIFT-PHOENIX & 0.003  & 0.003  & 0.004  & 0.023  & 0.025  & 0.038  \\
M8V    & DRIFT-PHOENIX & 0.003  & 0.003  & 0.003  & 0.024  & 0.025  & 0.035  \\
M9V    & DRIFT-PHOENIX & 0.003  & 0.003  & 0.003  & 0.024  & 0.026  & 0.037  \\
\hline
\multicolumn{8}{c}{Solar-like Spots and Faculae} \\
\hline
M0V    & PHOENIX       & 0.14   & 0.21   & 0.44   & 0.68   & 0.72   & 0.56   \\
M1V    & PHOENIX       & 0.16   & 0.23   & 0.34   & 0.69   & 0.72   & 0.65   \\
M2V    & PHOENIX       & 0.12   & 0.19   & 0.38   & 0.65   & 0.72   & 0.62   \\
M3V    & PHOENIX       & 0.10   & 0.15   & 0.26   & 0.58   & 0.69   & 0.71   \\
M4V    & PHOENIX       & 0.11   & 0.16   & 0.32   & 0.62   & 0.70   & 0.67   \\
M5V    & PHOENIX       & 0.06   & 0.16   & 0.33   & 0.45   & 0.70   & 0.66   \\
M5V    & DRIFT-PHOENIX & 0.02   & 0.14   & 0.21   & 0.15   & 0.68   & 0.72   \\
M6V    & DRIFT-PHOENIX & 0.01   & 0.06   & 0.25   & 0.09   & 0.44   & 0.71   \\
M7V    & DRIFT-PHOENIX & 0.01   & 0.10   & 0.26   & 0.08   & 0.60   & 0.71   \\
M8V    & DRIFT-PHOENIX & 0.01   & 0.10   & 0.26   & 0.08   & 0.58   & 0.71   \\
M9V    & DRIFT-PHOENIX & 0.01   & 0.05   & 0.24   & 0.08   & 0.39   & 0.72   \\
\enddata
\tablecomments{\added{Section~\ref{sec:methods_fspot} provides the definitions of the spot covering fractions $f_{spot, min}$, $f_{spot, mean}$, and $f_{spot, max}$, and the faculae covering fractions $f_{fac, min}$, $f_{fac, mean}$, and $f_{fac, max}$.}}
\end{deluxetable*}

Figure~\ref{fig:fspot_variability_faculae} shows the results of the variability modeling for the spots-and-faculae case. In contrast to the spots-only case, the addition of faculae leads to larger \added{I-band} variability full-amplitudes and a plateau in the relation for small spot covering fractions. As a result, a larger range of $f_{spot}$ corresponds to $A=0.01$ than in the spots-only case. Variability full-amplitudes begin to grow asymptotically again for $f_{spot}>0.1$ because the photospheres are nearly fully covered with faculae (due to the 10:1 faculae:spot ratio) and additional faculae do not contribute to the photospheric heterogeneity. Table~\ref{tab:filling_factors_spots_faculae} provides the range of $f_{spot}$ and $f_{fac}$ in the spots and faculae case for each spectral type.

For both spots-only and spots-and-faculae models, the variability modeling results show that extrapolations assuming a linear relation between the observed variability and spot covering fraction tend to underestimate the true spot covering fraction. Additionally, there is not a one-to-one relation between the observed variability and the spot covering fraction; instead, each observed variability full-amplitude corresponds to a range of covering fractions.

\subsection{Stellar contamination spectra}
\label{sec:spectra}

\begin{figure*}[!htbp]
\includegraphics[width=\linewidth]{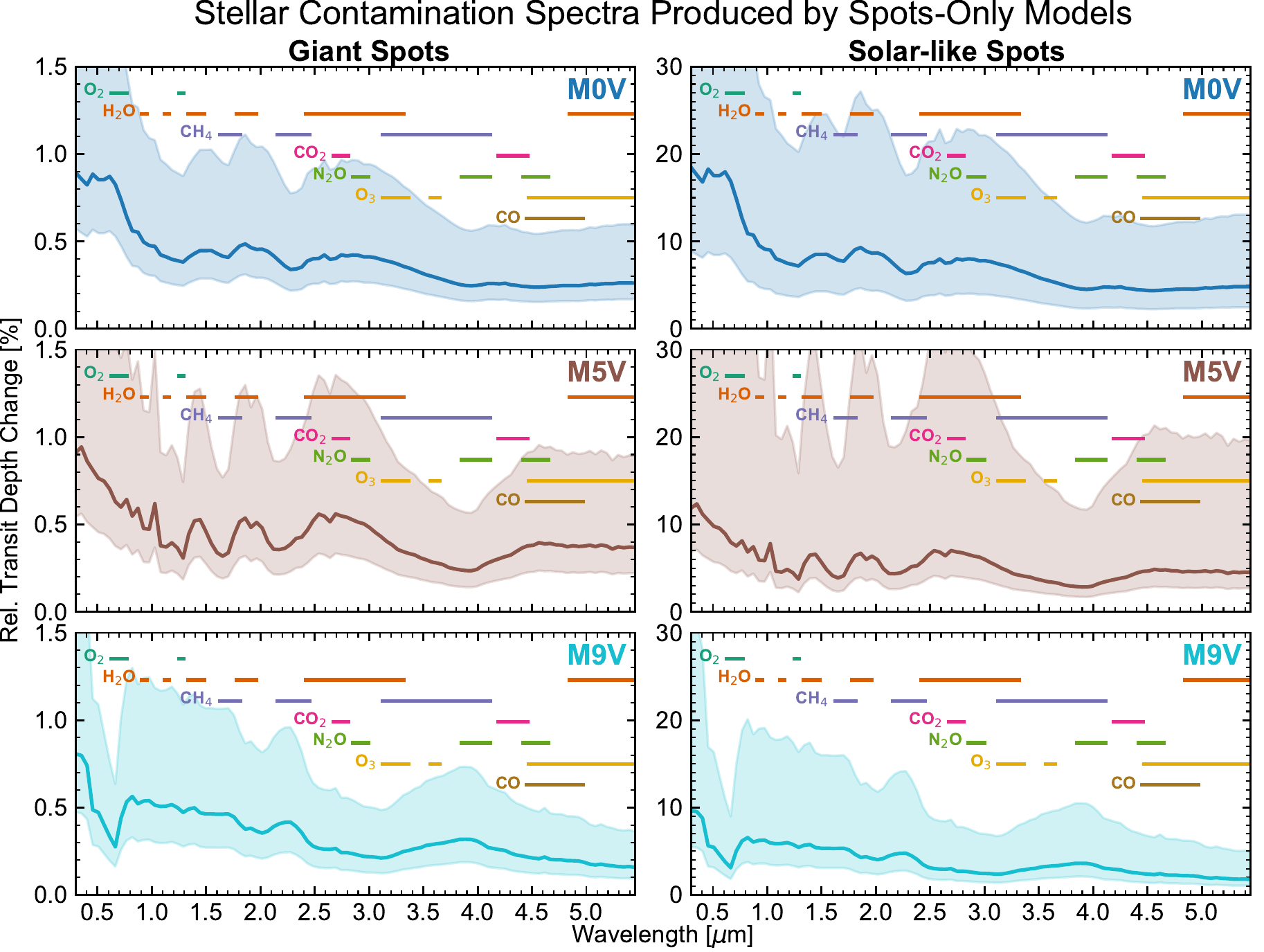}
\caption{Stellar contamination spectra produced by spots-only models. Contamination spectra for early, mid, and late M dwarf spectral types are shown. Models for giant (solar-like) spots are shown in the left (right) column. Solid lines give the contamination spectrum for the mean spot covering fraction consistent with a 1\% variability full-amplitude. The shaded regions illustrate the range of contamination spectra produced by spot covering fractions consistent with that same variability (see Table~\ref{tab:filling_factors_spots_only}). Overlapping wavelength bands for key exoplanetary atmospheric features are given. \label{fig:spots}}
\end{figure*}

Figure~\ref{fig:spots} shows the contamination spectra produced by the spots-only models for three representative spectral types. The shaded regions illustrate the range of contamination spectra possible due to the range of spot covering fractions allowed. Unocculted spots lead to an increase in transit depths across the full wavelength range studied, with the largest increases at the shortest wavelengths. Variations in transit depth due to differences in the molecular opacities between the photosphere and spots are also apparent. They are strongest for mid-M dwarfs and overlap  with many regions of interest for exoplanetary transmission features. We find that while the spectral type determines the specific features present, the overall scale of the contamination spectrum is largely independent of spectral type.  

Spot size, however, has a strong effect on the scale of the contamination spectra for a given covering fraction. Considering the mean spot covering fractions consistent with a 1\% variability full-amplitude, giant spots alter transit depths by an average of 0.4\% across \replaced{all}{0.3--5.5~$\micron$} wavelengths \replaced{and all}{for M dwarf} spectral types. Given the allowed range of spot covering fractions, however, the average change could be as low as 0.2\% or as high as 0.9\%. Solar-like spots, by contrast, produce much larger changes to transit depths due to the larger spot covering fractions present in this case. \replaced{The}{In this case, the} average change to transit depths \added{at these wavelengths} for the mean spot covering fraction is 5.0\% \deleted{for solar-like spots} and could be as low as 2.6\% or as high as 16\%, considering the allowed range of spot covering fractions.

\begin{figure*}[!htbp]
\includegraphics[width=\linewidth]{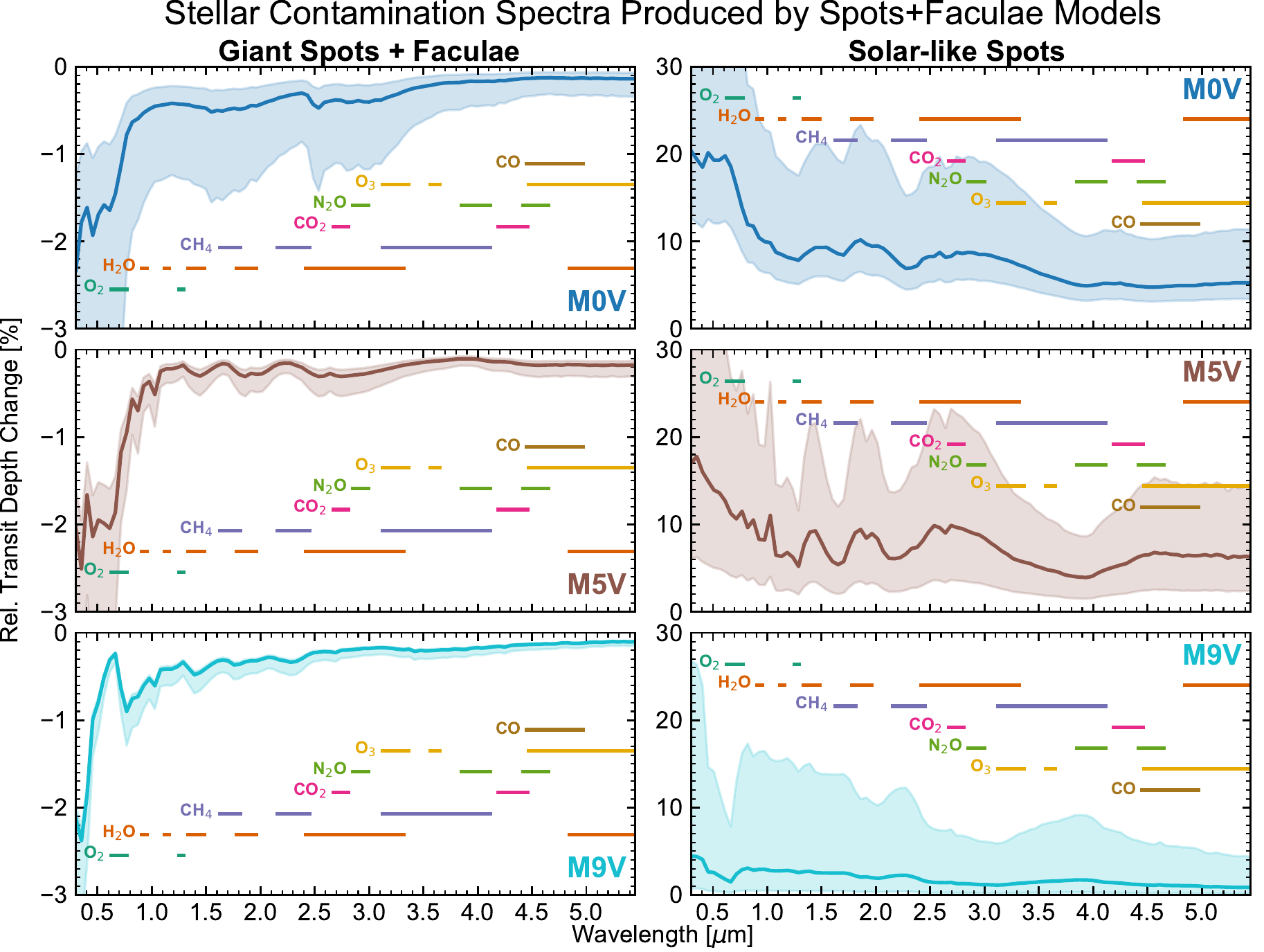}
\caption{Stellar contamination spectra produced by spots and faculae models. The figure elements are the same as those in Figure~\ref{fig:spots}. \label{fig:faculae}}
\end{figure*}

Figure~\ref{fig:faculae} shows contamination spectra produced by the spots and faculae variability models. In the giant spots case, the contamination spectra for all \added{M dwarf} spectral types are dominated \added{in the 0.3--5.5~$\micron$ wavelength range} by the facular contribution, which reduces transit depths at \replaced{all}{these} wavelengths. Comparing Tables~\ref{tab:filling_factors_spots_only} and ~\ref{tab:filling_factors_spots_faculae} shows that spot filling factors in all giant spots cases are very low, typically $\lesssim 1\%$, but they are on average $2.2 \times$ smaller in the giant spots and faculae case than in the giant spots-only case. Thus, a general finding is that the facular component can dominate the \added{visual and near-infrared} contamination spectra for low spot covering fractions. On average, in the giant spots and faculae case, transit depths are decreased by 0.4\% across \replaced{all wavelengths and spectral types}{this wavelength range for M dwarf spectral types}. Given the range of allowed spot- and faculae-covering fractions in these models, this average decrease could be as low as 0.3\% or high as 0.9\%. However, the contamination spectra of all \added{M dwarf} spectral types are more pronounced at shorter wavelengths, producing a mean decrease in transit depth of 2.4\% for \replaced{wavelengths $< 1~\mu$m}{0.3--1~$\micron$ wavelengths}. We note this result is consistent with the \replaced{optical}{visual} transmission spectrum of \object{GJ 1214b} \citep{Rackham2017}, which orbits a mid-M dwarf host star.

The case of solar-like spots and faculae presents a challenge to our original assumption of an immaculate transit chord. The facular coverages determined in these models are typically $> 50\%$ and range as high as $72\%$. If these facular coverages were indeed present, faculae would represent the dominant component of the stellar disk and likely the transit chord as well, precluding attempts to mask faculae crossings in light curves. Therefore, we assume in this case that the the faculae-covering fraction within the transit chord is roughly equivalent to that of the unocculted stellar disk and we calculate the contamination spectra shown in Figure~\ref{fig:faculae} without the contribution from a faculae heterogeneity. Nonetheless, the range of facular-covering fractions presented in Table~\ref{tab:filling_factors_spots_faculae}, particularly for late M dwarfs, shows that facular coverages are largely unconstrained and thus a wide range of facular contributions to stellar contamination spectra are possible for the solar-like spots case.

Regardless of the precise contribution from faculae, the addition of faculae to the heterogeneity model for the solar-like spots case generally increases the spot covering fractions that are consistent with \replaced{a}{an I-band} variability full-amplitude of 1\%  (Table~\ref{tab:filling_factors_spots_faculae}). On average, spot covering fractions in the solar-like spots and faculae case are $1.2 \times $ larger than those in the case of solar-like spots only, leading to an average increase in transit depth across \added{M dwarf} spectral types of 5.8\% \replaced{for all wavelengths studied}{in the 0.3--5.5~$\micron$ wavelength range}.

\section{Discussion} \label{sec:discussion}

We find that, for a range of host star spectral types, heterogeneous stellar photospheres strongly alter transmission spectra. In the following sections, we compare the spot and faculae covering fractions determined by our modeling efforts with empirical results, discuss the effect of stellar contamination on derived planetary parameters, and examine the example of the TRAPPIST-1 system in detail.

\subsection{Comparison with empirical results}
\label{sec:empirical_results}

Active region properties such as fractional areal coverage are difficult to ascertain in spatially unresolved observations of stars. However, estimates can be made based on the unique properties of chromospheric line formation in the specific case of M dwarf stars. In brief summary, the H$\alpha$ line appears only weakly in absorption in the cool photospheres of M dwarfs. However, as demonstrated by \citet{Cram1979}, the onset of chromospheric heating leads first to an increase in absorption strength of the line. With further non-radiative heating the line eventually attains a maximum in absorption equivalent width. Enhanced heating causes H$\alpha$ to become collisionally controlled and driven into emission as the defining observational characteristic of a dMe star.

Based on this general behavior of H$\alpha$ line formation in M dwarfs, \citet{Giampapa1985} discussed how the observed equivalent width of H$\alpha$ \textit{absorption} yields estimates of the \textit{minimum} area coverage of active regions. Since it is a chromospheric feature, its observed strength will be a function of its intrinsic absorption in plage and the total fractional area coverage of associated magnetically active facular regions. For specific M dwarf (non-dMe) stars with measurements available at that time, ranging in spectral type from M1.5 to M5.5, he deduced model-independent filling factors of facular regions exceeding 10--26\%. A more stringent inference of minimum filling factor can be obtained from calculations of the maximum H$\alpha$ absorption equivalent width attained in M dwarf model chromospheres. Based on these computations \citep{Cram1979}, \citet{Giampapa1985} found a range in the minimum active region filling factor of 31--67\% in the case of non-dMe stars characterized by $(R - I)_K$ = 0.9, corresponding to about T$_{eff}$ = 3500~K.  Therefore, the fractional area coverage of faculae of even relatively quiescent M dwarfs is widespread.

In the case of dMe stars, more intense chromospheric heating gives rise to H$\alpha$ emission which by itself does not lead to a direct estimate of facular area coverage. However, there is direct observational evidence for widespread, multi-kilogauss magnetic fields outside of spots in active M dwarfs (i.e., dMe stars) at high filling factors in excess of 50\% based on the analysis of magnetically sensitive photospheric features \citep{Saar1985, Reiners2009}. Furthermore, the absence of any reported rotational modulation of H$\alpha$ emission in dMe stars is consistent with the occurrence of an azimuthally symmetric, spatially widespread presence of facular regions on their surfaces.  

In addition to active regions characterized by bright chromospheric emission or, in the case of quiescent M dwarfs, chromospheric H$\alpha$ absorption, the occurrence of periodic or quasi-periodic light curve modulations in photometric bandpasses, indicative of the presence of cool spots, has an extensive record of observations. In a study of a compilation of extensive datasets, \citet{Newton2016, Newton2017} measured photometric semi-amplitudes in the range of roughly 0.5--4\% \citep[Fig.~9]{Newton2017}. We can therefore infer minimum starspot filling factors $f_{spot} \sim$ 1--8\% of the visible stellar disk in the case of completely black spots. Estimates of starspot temperatures provide a further refinement of the minimum spot filling factor. In an investigation of the magnetic properties of spots on the active, early M dwarf AU Mic (dM1e), \citet{Berdyugina2011} utilized polarimetric observations of temperature- and magnetically-sensitive diatomic molecules to find spot temperatures $\sim$ 500--700~K cooler than the surrounding photosphere, and with associated magnetic field strengths as high as 5.3 kG.  With an effective temperature of 3775~K \citep{Pagano2000}, the ratio of spot-to-stellar effective temperature is in the range of 0.81--0.87. In general, \citet{Afram2015} find for M dwarfs that spot models tend to have a spot-to-photosphere temperature ratio of 0.86. Based on these measurements, for a semi-amplitude of $\sim$ 2\% with a range of T$_{spot}$/T$_{eff}$ of $\sim$ 0.8 to 0.9 we estimate spot filling factors in the range of $\sim$ 7--12\%. Given that the photometric modulation depends on both contrast and departures from a longitudinally symmetric surface distribution of spots, this spot filling factor estimate should be regarded as a minimum value since the observed light curve modulation could arise from departures from axial symmetry in an otherwise widespread and uniform distribution of small spots. In fact, \citet{Jackson2013} demonstrate that small-amplitude light curve modulation can arise from the combined effect of a random distribution of small spots characterized by a uniform size (i.e., scale length) and a plausible temperature contrast with the surrounding photosphere of 0.7. As an illustration, their Monte Carlo simulations of light-curve amplitudes due to a large number of randomly distributed spots with, say, a characteristic length scale of 3.5 degrees yield a mean amplitude of 1\% with a filling factor of $\sim$ 30\% \citep[Fig.~4]{Jackson2013} in qualitative agreement with our variability modeling. Thus, in this context, we conclude that high filling factors of spots and faculae, such as those determined in this work, are a plausible interpretation of the observed light curve modulations seen in M dwarfs.

\subsection{Stellar contamination mimicking and masking exoplanetary features}
\label{sec:features}

As shown in Figures~\ref{fig:spots} and \ref{fig:faculae}, stellar heterogeneity can introduce significant spectral features in \added{visual and near-infrared} transmission spectra. These deviations result from a difference in flux between the immaculate photosphere and heterogeneities such as spots and faculae (Equation~\ref{eq:CS_sf}). Unocculted spots introduce positive features in transmission spectra that may be mistaken for evidence of absorption or scattering in the exoplanet atmosphere. By contrast, unocculted faculae introduce negative features, which can mask genuine spectral features originating in the exoplanet atmosphere.

\begin{figure*}[!htbp]
\includegraphics[width=\linewidth]{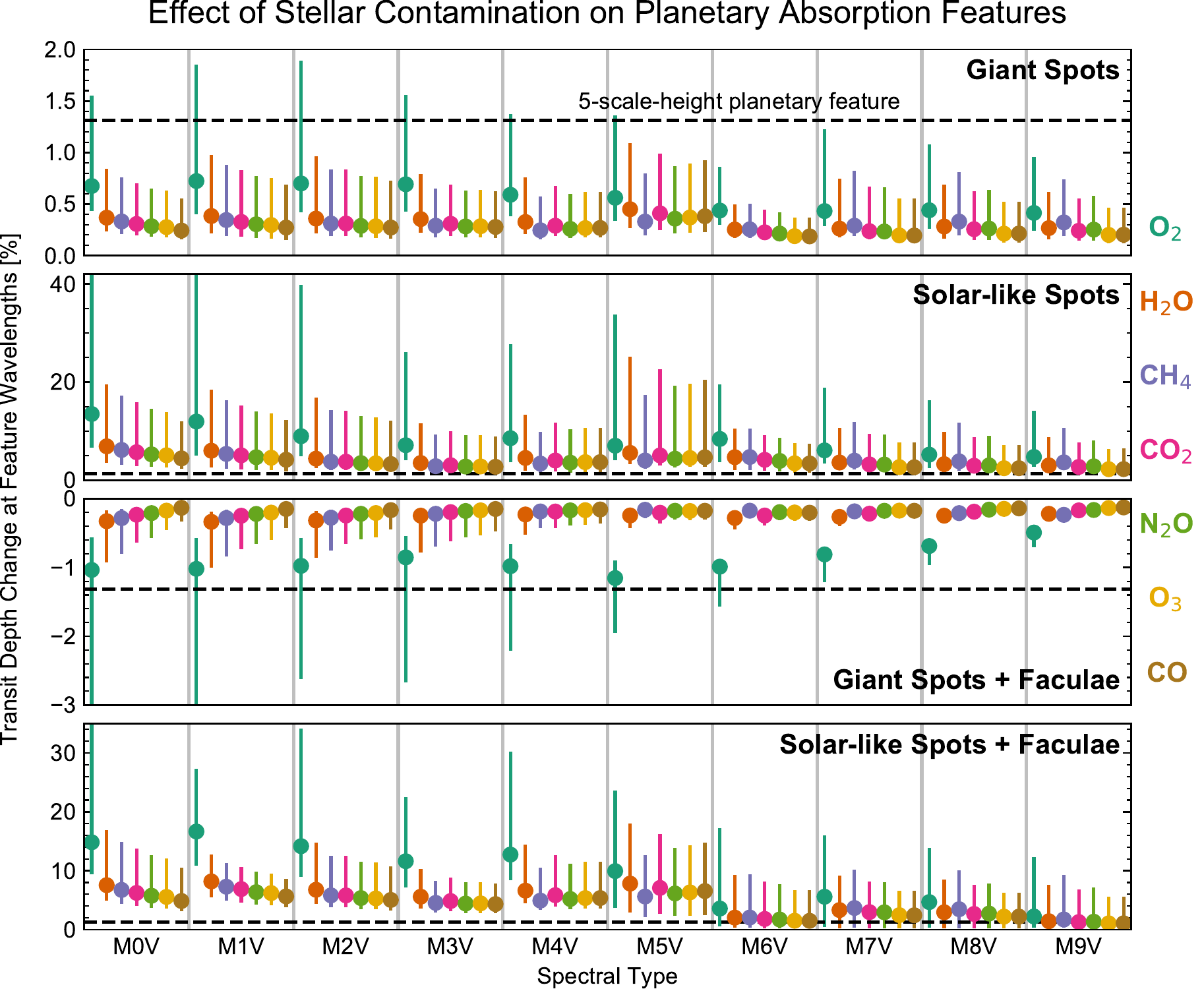}
\caption{Effect of stellar contamination at wavelengths of planetary absorption features. The change in transit depth integrated over wavelengths of interest for molecular species is shown as a function of spectral type. From top to bottom, the panels illustrate results for four cases: giant spots, solar-like spots, giant spots and faculae, and solar-like spots and faculae. Points correspond to the mean stellar contamination signal and error bars to the dispersion in the strength of the signal illustrated by the shaded regions in Figures~\ref{fig:spots} and \ref{fig:faculae}. Horizontal dashed lines show the expected transit depth change for a feature covering five scale heights in an Earth-like atmosphere, which is independent of stellar spectral type. \label{fig:features}}
\end{figure*}

Figure~\ref{fig:features} illustrates the effect of stellar contamination at \added{visual and near-infrared} wavelengths of interest for molecular features in exoplanetary atmospheres. In general, stellar contamination increases transit depths and may mimic exoplanetary features, with the exception in the case of giant spots and faculae, in which the contribution from faculae dominates and may mask exoplanetary features. Transit depth changes are largest and therefore most problematic for:  1) solar-like spots cases, 2) earlier spectral types, and 3) molecules with absorption bands at relatively short wavelengths, such as $O_{2}$.

As a comparison for scale, the dashed horizontal line in Figure~\ref{fig:features} indicates the 1.3\% change in transit depth expected for a transiting Earth-twin (atmospheric mean molecular weight $\mu = 28.97$~amu; equilibrium temperature $T_{eq} = 288$~K) due to an atmospheric feature covering five pressure scale heights. In the giant spots cases, stellar contamination can alter transit depths at overlapping wavelengths by a non-negligible fraction of this amount, the effect of which is to boost the apparent size of features in the spots-only case and to weaken them in the giant spots and faculae case. Therefore, large unocculted spots can lead to a range of erroneous interpretations of transmission spectra: molecular abundances may appear enhanced or depleted and the presence of a obscuring haze layer can be masked or mimicked.

By contrast, solar-like spots cases increase transit depths by much more than an expected exoplanetary feature. For early spectral types, the increase in transit depth can be more than \textit{ten times} that expected for planetary features. Such a strong feature could be easily identified as stellar in origin, though the scale of the feature, combined with the wide range of stellar contributions possible, would limit the accuracy of any determination of the underlying planetary feature. Later spectral types present a more pathological circumstance, as the magnitude of the transit depth change due to stellar contamination is comparable to that of a planetary atmospheric feature, allowing the signals to be easily mistaken. Therefore, constraining spot sizes and, by extension, spot covering fractions, will be essential for investigations of atmospheric features in transmission spectra of low-mass exoplanets around late M dwarfs, such as the TRAPPIST-1 system. We consider this topic in more detail in Section~\ref{sec:TRAPPIST-1} below.

\added{
\subsection{Comparison to observational precisions}
\label{sec:observational_precisions}

Thus far we have discussed the effect of stellar heterogeneity on transmission spectra in terms of relative transit depth changes. This approach is useful because the multiplicative change in transit depth produced by unocculted spots and faculae can be applied to any planetary transmission spectrum. However, it is also informative to convert these relative transit depth changes to absolute ones that can be compared to observational precisions afforded by current and near-future facilities.

To do this, we must adopt parameters for the transiting exoplanet. At a minimum, these are the planetary radius and the expected scale of a planetary atmospheric feature. We consider two end-member cases for this comparison: a hot Neptune and an Earth-twin. For the hot Neptune case, we adopt the parameters of GJ~436b \citep{Butler2004, Gillon2007}, which is one of the largest and hottest planets known to transit an M dwarf and represents a more readily observable exoplanet atmosphere. Its near-infrared transmission spectrum \citep{Knutson2014} displays a weighted mean transit depth of $\overline{D} = 0.70\%$ and a 200-ppm variation in transit depth \citep{Knutson2014}, which has been interpreted as an H$_{2}$O absorption feature covering $0.46 \pm 0.25$ atmospheric scale heights \citep{Crossfield2017}. This feature produces a relative transit depth change of 
\begin{equation}
\epsilon_{p} = \frac{\Delta D}{D}
             = \frac{2.0 \times 10^{-4}}{7.0 \times 10^{-3}}
             = 2.9\%.
\end{equation}
We adopt this observed relative transit depth change and the best-fit planetary radius of GJ~436b \citep[$R_{p} = 3.95~R_{\earth}$;][]{Gillon2007} for the hot Neptune case. The other end-member case we consider is that of a transiting Earth-twin ($R_{p} = 1.0~R_{\earth}$; $\mu = 28.97$~amu; $T_{eq} = 288$~K), which produces a relative transit depth change of $\epsilon_{p} = 1.3\%$ for a 5-scale-height planetary feature and represents a larger observational challenge.

For each of these planetary cases, we calculate transit depths $D$ and absolute transit depth changes due to planetary atmospheric features $\Delta D_{p}$ as a function of M dwarf spectral type using the stellar radii provided in Table~\ref{tab:stars}. We also calculate the mean absolute transit depth change due to stellar heterogeneity $\Delta D_{s}$ at wavelengths of interest for H$_{2}$O absorption (as discussed in Section~\ref{sec:features}) for each of the four heterogeneity cases we consider. The results are summarized in Tables~\ref{tab:hotneptune} and \ref{tab:earthtwin} for the hot Neptune and Earth-twin cases, respectively. Comparing these values of $\Delta D_{p}$ and $\Delta D_{s}$ suggests that for hot Neptunes variations in transit depth due to the planetary atmosphere are roughly an order of magnitude larger than those due to stellar heterogeneity for giant spot cases. However, stellar signals are larger than planetary signals for all solar-like spot cases, with the largest difference between the signals for earlier spectral types. For Earth-twins transiting M dwarfs, in the case of giant spots, planetary signals are at least three times larger than stellar ones for all M dwarf spectral types. However, as with the hot Neptune case, stellar signals due to solar-like spots are larger than planetary signals for all M dwarf spectral types.

The question remains, however, of how these planetary and stellar signals compare to observational precisions. The most precise transmission spectrum obtained with \textit{HST/WFC3} to date contains a typical uncertainty of 30 ppm on the transit depth in each wavelength channel \citep{Kreidberg2014}. This is similar to the noise floors for James Webb Space Telescope (\textit{JWST}) instruments adopted by \citet{Greene2016} for wavelength ranges of interest in this study: 20 ppm for NIRISS SOSS ($\lambda$ = 1--2.5~$\micron$) and 30 ppm for NIRCam grism ($\lambda$ = 2.5--5.0~$\micron$). We adopt 30 ppm as a fiducial noise floor for both \textit{HST} and \textit{JWST} observations. Considering this detection threshold, our results suggest that planetary atmospheric features are detectable for hot Neptunes orbiting M dwarfs of all spectral types (Table~\ref{tab:hotneptune}). The effects of unocculted giant spots and facular regions are detectable for host stars with spectral types of roughly M3V and later, while in the more problematic case of solar-like spots, the effects of unocculted spots and faculae are detectable for all M dwarf spectral types. In the case of a transiting Earth-twin (Table~\ref{tab:earthtwin}), we estimate that planetary atmosphere features are detectable for spectral types M6V and later. The effects of unocculted giant spots and facular regions may alter the strength of these features by $\sim 20\%$ but will only independently reach the detection threshold of 30 ppm for M9V host stars. However, in the case of solar-like spots, we estimate that the effects of unocculted spots and faculae may be apparent in observations of spectral types as early as M3V or M4V.

In summary, we find that the most precise existing \textit{HST/WFC3} G141 transmission spectra as well as upcoming \text{JWST} transmission spectra of small planets around M dwarfs can be significantly influenced by stellar contamination; any analysis of such spectra should consider the possible range of systematic contamination due to the transit light source effect.

\begin{deluxetable*}{ccccccc}[!tbp]
\tabletypesize{\small}
\tablecaption{Transit depths and absolute transit depth changes for a representative hot Neptune by spectral type \label{tab:hotneptune}}
\tablehead{
		   \colhead{Sp. Type}            &
		   \colhead{$D$}                 &
           \colhead{$\Delta D_{p}$}          &
           \multicolumn{4}{c}{$\Delta D_{s}$ by Heterogeneity Case}  \\
           \cline{4-7} 
           \colhead{}                    &
		   \colhead{}                    &
           \colhead{}                    &
           \colhead{Giant Spots}         &
		   \colhead{Solar-like Spots}    &
		   \colhead{Giant Spots + Faculae}         &
           \colhead{Solar-like Spots + Faculae}    \\
           \colhead{}                         &
           \colhead{(ppm)}                    &
           \colhead{(ppm)}                    &
		   \colhead{(ppm)}                    &
		   \colhead{(ppm)}                    &
		   \colhead{(ppm)}                    &
           \colhead{(ppm)}      
		  }
\startdata
M0V      & 3400     & 99       & 13            & \bf{240}      & -11           & \bf{260}      \\
M1V      & 5400     & 160      & 21            & \bf{330}      & -18           & \bf{450}      \\
M2V      & 6800     & 200      & 24            & \bf{300}      & -21           & \bf{460}      \\
M3V      & 8600     & 250      & 30            & \bf{300}      & -21           & \bf{480}      \\
M4V      & 19000    & 560      & 63            & \bf{880}      & -44           & \bf{1300}     \\
M5V      & 33000    & 950      & 100           & \bf{1100}     & -85           & \bf{1900}     \\
M6V      & 58000    & 1700     & 150           & \bf{2700}     & -160          & 1200          \\
M7V      & 91000    & 2600     & 240           & \bf{3300}     & -240          & \bf{3000}     \\
M8V      & 110000   & 3100     & 300           & \bf{3600}     & -260          & \bf{3200}     \\
M9V      & 200000   & 5900     & 550           & \bf{6200}     & -450          & 2900          \\
\enddata
\tablecomments{\added{Listed are the planetary transit depth $D$, transit depth change due to planetary atmospheric features $\Delta D_{p}$, and, for the four heterogeneity cases we consider, the transit depth change due to stellar heterogeneity $\Delta D_{s}$ (shown in bold for cases in which the stellar transit depth change is larger than that due to planetary atmospheric features). The scale of the stellar signal is smaller than that of a hot Neptune atmospheric feature for all giant spots cases, but it is generally larger than the planetary signal for cases with solar-like spots.}}
\end{deluxetable*}

\begin{deluxetable*}{ccccccc}[!tbp]
\tabletypesize{\small}
\tablecaption{Transit depths and absolute transit depth changes for a transiting Earth-twin by spectral type \label{tab:earthtwin}}
\tablehead{
		   \colhead{Sp. Type}            &
		   \colhead{$D$}                 &
           \colhead{$\Delta D_{p}$}          &
           \multicolumn{4}{c}{$\Delta D_{s}$ by Heterogeneity Case}  \\
           \cline{4-7} 
           \colhead{}                    &
		   \colhead{}                    &
           \colhead{}                    &
           \colhead{Giant Spots}         &
		   \colhead{Solar-like Spots}    &
		   \colhead{Giant Spots + Faculae}         &
           \colhead{Solar-like Spots + Faculae}    \\
           \colhead{}                         &
           \colhead{(ppm)}                    &
           \colhead{(ppm)}                    &
		   \colhead{(ppm)}                    &
		   \colhead{(ppm)}                    &
		   \colhead{(ppm)}                    &
           \colhead{(ppm)}      
		  }
\startdata
M0V      & 220      & 2.8      & 0.8      & \bf{15}       & -0.7     & \bf{16}       \\
M1V      & 350      & 4.5      & 1.3      & \bf{21}       & -1.2     & \bf{29}       \\
M2V      & 430      & 5.6      & 1.6      & \bf{19}       & -1.4     & \bf{29}       \\
M3V      & 550      & 7.2      & 2.0      & \bf{19}       & -1.4     & \bf{31}       \\
M4V      & 1200     & 16       & 4.0      & \bf{56}       & -2.8     & \bf{82}       \\
M5V      & 2100     & 27       & 6.5      & \bf{73}       & -5.4     & \bf{120}      \\
M6V      & 3700     & 48       & 9.4      & \bf{180}      & -10      & \bf{76}       \\
M7V      & 5800     & 76       & 15       & \bf{210}      & -15      & \bf{190}      \\
M8V      & 6900     & 90       & 20       & \bf{230}      & -17      & \bf{210}      \\
M9V      & 13000    & 170      & 35       & \bf{400}      & -29      & \bf{190}      \\
\enddata
\tablecomments{\added{Listed are the planetary transit depth $D$, transit depth change due planetary atmospheric features $\Delta D_{p}$, and, for the four heterogeneity cases we consider, the transit depth change due to stellar heterogeneity $\Delta D_{s}$ (shown in bold for cases in which the stellar transit depth change is larger than that due to planetary atmospheric features). The scale of the stellar signal is smaller than that of an Earth-twin atmospheric feature for all giant spots cases, but it is larger than the planetary signal for all cases with solar-like spots.}}
\end{deluxetable*}

}

\subsection{Systematic errors in density measurements} \label{sec:densities}

As the fundamental effect of starspots in this context is to influence the apparent radius of the planet relative that of its star, we also explore here the impact of starspots on planet density measurements. Transiting planet surveys utilize visual and near-infrared bandpasses for discovery efforts \citep[e.g.,][]{Nutzman2008, Jehin2011, Gillon2011}. With the exception of \textit{Kepler}, which utilizes a broad, unfiltered \replaced{optical}{visual} bandpass, these surveys typically favor redder wavelengths such as I-band in order to minimize the \replaced{stellar contribution}{contribution of stellar variability} to measurements. The bulk density of an exoplanet calculated from discovery transits can provide a clue as to their volatile content, making it one of the primary factors affecting whether a planet is selected for follow-up observations. Furthermore, the unfiltered \textit{Kepler} photometry often remains the most precise transit depth measurement for most small planets, and therefore its accuracy affects inferences made about individual planets as well as ensembles of planets.

\begin{figure*}[t]
\includegraphics[width=\linewidth]{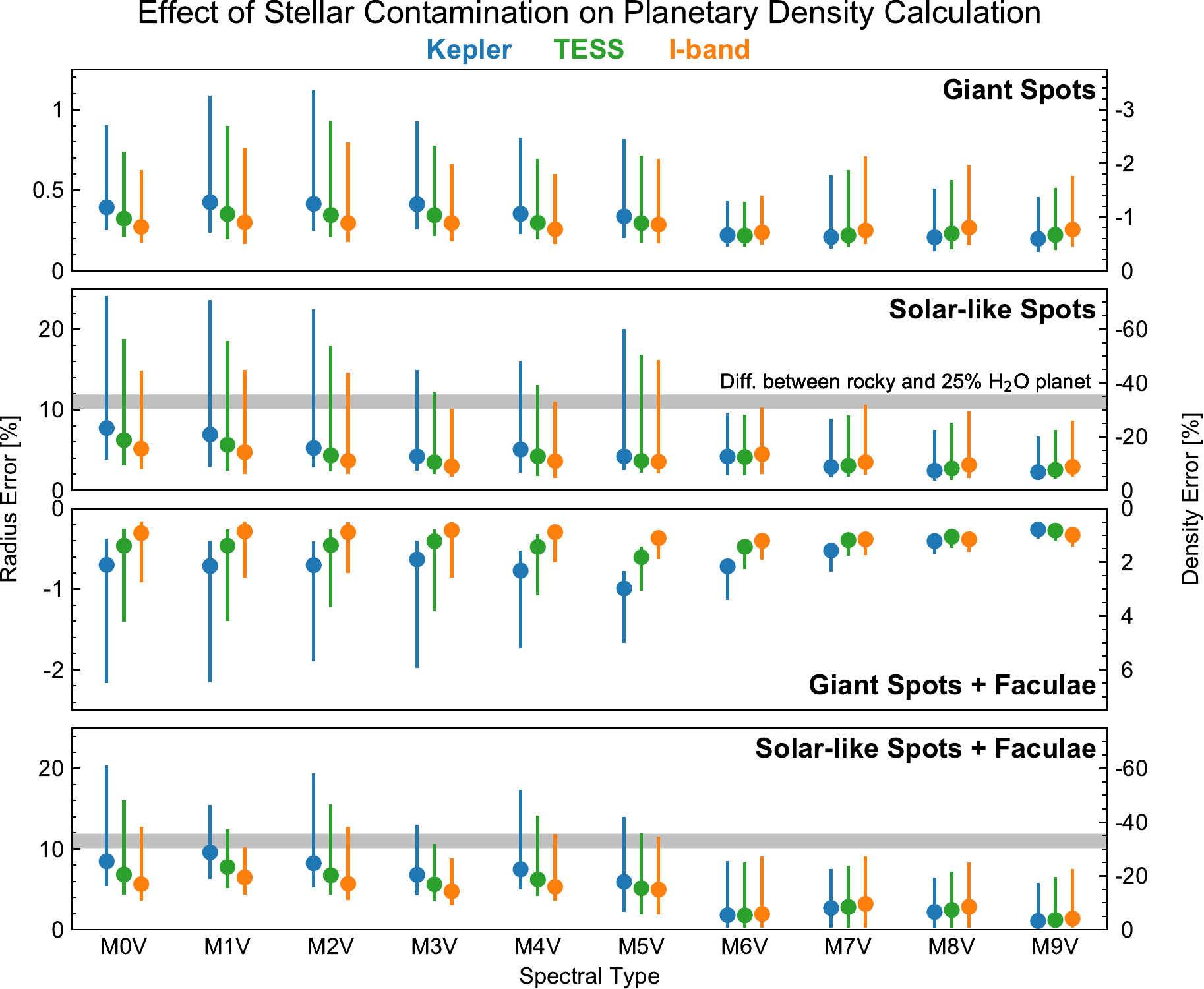}
\caption{Effect of stellar contamination on planetary density calculation. The systematic radius error is shown as a function of spectral type. The corresponding density error is shown at right. Note the difference in sign. The panels illustrate results from four cases of heterogeneities as shown in Figure~\ref{fig:features}. Colors correspond to the \textit{Kepler} (blue), \textit{TESS} (green), and I-band (orange) photometric bands. The mean radius error is shown as a point and error bars correspond to the dispersion in the strength of the contamination signal (see Figures~\ref{fig:spots} and \ref{fig:faculae}). The gray shaded region illustrates the relative difference in radius between an entirely rocky exoplanet and one with a significant volatile content for a range of masses between 0.125 and 32 Earth masses (see text). \label{fig:density}}
\end{figure*}

Our work here, however, shows that the stellar contributions can be larger than previously suspected and may still contribute significantly even at red \replaced{optical}{visual} and near-infrared wavelengths, given the large range of stellar heterogeneity levels that may correspond to an observed rotational variability level. To determine the scale of the contribution, we calculated the range of radius and, by extension, density errors by integrating the contamination spectra presented in Section~\ref{sec:spectra} over the Bessel I-band filter bandpass, which is similar to bandpasses used by ground-based transit surveys such as MEarth \citep{Nutzman2008} and TRAPPIST \citep{Gillon2011, Jehin2011}, and the \textit{Kepler}\footnote{\url{https://archive.stsci.edu/kepler/fpc.html}} and \textit{TESS} \citep{Ricker2015, Sullivan2015} spectral responses.

The resulting systematic radius and density errors are shown in Figure~\ref{fig:density}. Positive radius errors, which indicate the apparent planetary radius is larger than the true planetary radius, correspond to negative density errors, which indicate that the apparent density is smaller than the true bulk density. We note that this systematic density error provides only a minimum estimate of the total error in planetary bulk density, to which the observational uncertainty in radius and any error in mass estimate must be also be added. 

As with the investigation into false planetary absorption features, we find the primary determinant of the systematic density error to be the spot size. Small spots cases are the most problematic, with an average density error of 13\% across all spectral types and bandpasses \added{that we considered} and a maximum of 29\% for M1 dwarfs in the \textit{Kepler} bandpass. In the giant spots cases, by contrast, density errors are 1\% on average and the largest are 3\%. Aside from spot size, spectral type is the second-largest determinant of the systematic error. Earlier \added{M dwarf} spectral types tend to produce density errors that are larger on average. Additionally, they demonstrate a larger dispersion in possible errors, as indicated by the error bars, given the wider range of spot and faculae covering fractions corresponding to an observed variability (see Tables~\ref{tab:filling_factors_spots_only} and \ref{tab:filling_factors_spots_faculae}). Finally, we find that the systematic errors are generally larger for bluer observational bandpasses, though the effect of the bandpass is smaller than that of either the spot size or host star spectral type.

To place the systematic radius errors in context, the difference in radius between an entirely rocky composition and a planet with 25\% $H_{2}O$ by mass for planets with masses between 0.125 and 32 Earth masses is 10–-12\% \citep[Table~2]{Zeng2016}, illustrated by the shaded regions in Figure~\ref{fig:density}. In terms of an observational comparison, we note that the 1-sigma uncertainty on the radius of LHS~1140b is 7.0\% \citep{Dittmann2017}, giving it one of the best constrained densities of a sub-Neptune exoplanet to date. Therefore, while the effect of unocculted giant spots is easily hidden within observational uncertainties, unocculted solar-like spots can introduce a systematic error in radius that is comparable to the total error combined from the observational uncertainties in the radial velocity and transit photometry measurements and can significantly alter interpretations of the volatile content of an exoplanet. We note that this applies to M dwarf planets that will be discovered by \text{TESS}, for which we predict sunspot-like spots will lead on average to systematic radii overestimates of 4.3\% and density underestimates of 13\%.

\subsection{Application to TRAPPIST-1 system}
\label{sec:TRAPPIST-1}

We utilized the approach detailed in Section~\ref{sec:methods} to place constraints on spot and faculae covering fractions for TRAPPIST-1 and their effects on observations of the TRAPPIST-1 planets. We opted to explore the case of the TRAPPIST-1 planets because currently they represent the best examples of habitable zone, possibly rocky planets and these planets are also likely to be primary targets for in-depth \replaced{James Webb Space Telescope}{\textit{JWST}} transit spectroscopy. 

\subsubsection{Spot and faculae covering fractions}
\label{sec:TRAPPIST-1_covering_fractions}

We adopted the stellar parameters for TRAPPIST-1 from \citet{Gillon2017}, including stellar mass, radius, effective temperature, and metallicity. With these parameters, we \replaced{generated spectra using}{interpolated spectra covering 0.3--5.5~$\micron$ from} the DRIFT-PHOENIX model grid. We set the photosphere temperature to the effective temperature and calculated the spot and faculae temperatures using the relations outlined in Section~\ref{sec:component_spectra}.

We integrated these spectra over the Bessel I-band response, similar to the I+z bandpass utilized by the TRAPPIST survey \citep{Gillon2016, Gillon2017}, to simulate the contributions of these spectral components to photometric observations. From visual inspection of Extended Data Figure~5 of \citet{Gillon2016}, we estimate the variability full-amplitude of TRAPPIST-1 to be 1~\%.

Long-baseline monitoring of TRAPPIST-1 using the Spitzer Space Telescope, covering 35 transits, shows no definitive evidence of spot crossings \citep{Gillon2017}.
\added{Likewise, no spot crossings are apparent in existing \textit{HST/WFC3} transit observations of TRAPPIST-1b and TRAPPIST-1c covering 1.1--1.7~$\micron$ with the G141 grism \citep{deWit2016}, though the precision of G141 observations has allowed for the detection of large spot crossing events in other M dwarf transit observations \citep{Kreidberg2014}.}
\replaced{The}{Therefore, we assume that the} presence of very large spots like those detected through molecular spectropolarimetry on active M dwarfs \citep{Berdyugina2011}, each covering 0.5\% of the projected stellar disk and comparable in size to the TRAPPIST-1 planets, is unlikely. Accordingly, we adopt a spot size of $2 \degr$ (400~ppm or 0.04\% of the projected stellar disk), typical of large spot groups on the Sun \citep{Mandal2017}. Given the radius of TRAPPIST-1 \citep[$R=0.117 \pm 0.0036$;][]{Gillon2017}, this translates to spots with a radius of $R_{spot} = 1.63 \pm 0.50 \times 10^{3}$~km.

\begin{figure*}[htbp!]
\epsscale{1.15}
\includegraphics[width=\linewidth]{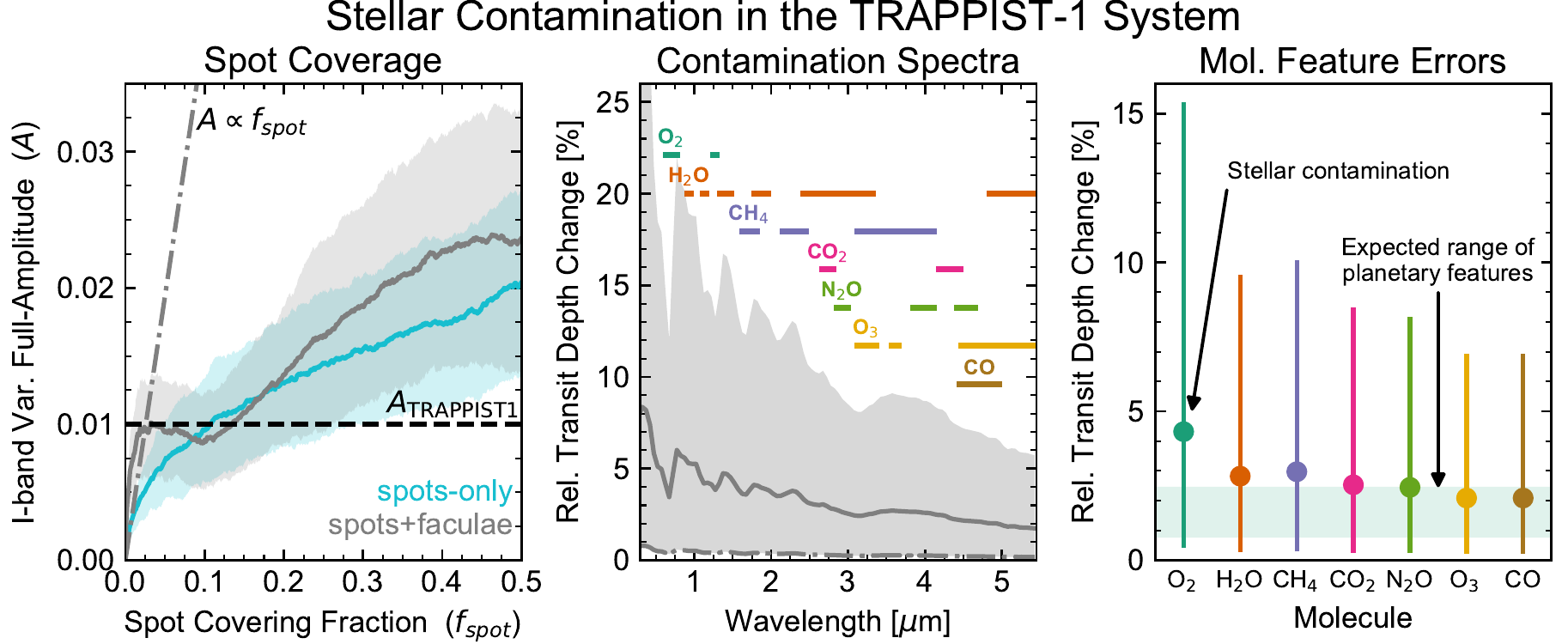}
\caption{Summary of the effect of stellar heterogeneity in the TRAPPIST-1 system. The left panel shows I-band variability full-amplitudes as a function of spot covering fraction for spots-only (teal) and spots and faculae (gray) cases. Results from solar-like spots models are shown. The plot elements are the same as those in Figure~\ref{fig:fspot_variability}. The central panel shows the contamination spectra produced by spot covering fractions consistent with the observed variability of TRAPPIST-1. The plot elements are the same as those in Figure~\ref{fig:spots}. The right panel shows the relative transit depth change at wavelengths of interest for key molecular features in exoplanet atmospheres. The shaded region illustrates the range of relative transit depth changes of interest for atmospheric features from the TRAPPIST-1 planets, assuming a 5 scale height feature in an Earth-like atmosphere \label{fig:TRAPPIST1}}
\end{figure*}

The left panel of Figure~\ref{fig:TRAPPIST1} displays the results of our variability modeling for both spots-only and spots and faculae cases. For both cases, we find a large range of spot covering fractions is consistent with the observed variability full-amplitude. We find 
\replaced{$f_{spot} = 10^{+19}_{-5}\%$}
{$f_{spot} = 11^{+18}_{-6}\%$} 
for the spots-only case, in which the quoted value is the mean spot covering fraction and the uncertainty refers to 68\% dispersion in model results. For the \replaced{spots+faculae}{spots and faculae} case, we find
\replaced{$f_{spot} = 12^{+10}_{-11}\%$ and $f_{fac} = 65^{+7}_{-55}\%$}
{$f_{spot} = 8^{+18}_{-7}\%$ and $f_{fac} = 54^{+16}_{-46}\%$}.
By contrast, under the assumption that the distribution of photospheric heterogeneities is fully asymmetric, in which case the variability amplitude would scale linearly with the spot covering fraction with a slope determined by the spot and photosphere I-band fluxes, one would infer a spot covering fraction of 1.0\%. We conclude that spot and faculae covering fractions for TRAPPIST-1 are largely unconstrained by the observed photometric variability amplitude, though the allowed range of fractions is typically much larger than one would infer by assuming a linear relation between spot covering fraction and variability amplitude.

\subsubsection{Stellar contamination in transmission spectra}
\label{sec:TRAPPIST-1_stellar_contamination}

The \added{visual and near-infrared} contamination spectra produced by the wide range of spots we infer for TRAPPIST-1 are shown in the center panel of Figure~\ref{fig:TRAPPIST1}. The stellar contamination is a multiplicative effect and independent of the planetary transmission spectrum (Equation~\ref{eq:CS}). Thus, these relative transit depth changes are applicable to observations of each exoplanet in the TRAPPIST-1 system. As with the solar-like spot models in Figure~\ref{fig:faculae}, contributions from unocculted faculae are not included, given that faculae covering fractions in our models are typically $> 50\%$ and are therefore likely to be distributed homogeneously throughout the stellar disk. However, we calculate the contribution of unocculted spots using the spot covering fractions determined by the \replaced{spots+faculae}{spots and faculae} models, which are generally smaller than those from the spots-only models, to exploit the increased realism of the spots and faculae models. The solid line shows the contamination spectrum corresponding to the mean spot covering fraction and the shaded region the range of spectra for the 68\% confidence interval on $f_{spot}$. Compared to the contamination spectrum predicted by the fully asymmetric assumption, shown as a dash-dotted line, we find the level of stellar contamination to be roughly an order of magnitude larger.

Integrating over wavelengths of interest for key planetary molecular features, we illustrate the practical impact on observations in the right panel of Figure~\ref{fig:TRAPPIST1}. For all molecules considered, we find the mean relative transit depth change to be \added{comparable to or} larger than that produced by an exoplanetary atmospheric feature. The shaded region in this panel illustrates the range of feature scales for the six innermost TRAPPIST-1 planets, calculated using the planetary parameters from \citet{Gillon2017} and assuming an Earth-like atmospheric mean molecular weight ($\mu=28.97$~amu). They range from 0.8\% for TRAPPIST-1g to 2.5\% for TRAPPIST-1d. \added{Given the transit depths of the TRAPPIST-1 planets, these correspond to absolute transit depth changes of 59~ppm (TRAPPIST-1g) to 168~ppm (TRAPPIST-1b). Assuming the same 30~ppm detection threshold as in Section~\ref{sec:observational_precisions}, atmospheric features with these scales are, in principle, detectable with both \textit{HST} and \textit{JWST}.} \replaced{By contrast}{In contrast to the planetary atmospheric signal}, mean molecular feature errors produced by unocculted spots range from \replaced{3.1\%}{2.1\%} for CO to \replaced{7.4\%}{4.3\%} for O$_{2}$. \added{For the planet with the shallowest transit depth in this set, TRAPPIST-1d \citep[$D=0.367 \pm 0.017$~\%;][]{Gillon2017}, a 2.1\% relative change in transit depth corresponds to an absolute transit depth change of 77~ppm, which illustrates that the range of uncertainties we predict for all molecular features for the TRAPPIST-1 planets are above the fiducial 30~ppm noise floor. In other words, we predict stellar heterogeneity will significantly impact the interpretation of high-precision TRAPPIST-1 transmission spectra.} Errors are most pronounced for molecules with absorption bands at shorter wavelengths. Large uncertainties exist for each molecular feature error due to the wide range of spot covering fractions and accompanying contamination spectra possible.

\begin{deluxetable}{ccc}[!tbp]
\tabletypesize{\small}
\tablecaption{Revised densities for the TRAPPIST-1 planets \label{tab:TRAPPIST1}}
\tablehead{
    \colhead{Planet} & \colhead{Density from} & \colhead{Revised Density} \\
    \colhead{} & \colhead{ \citet{Gillon2017}} & \colhead{from This Work} \\
    \colhead{} & \colhead{($\rho_{\earth}$\tablenotemark{a})} & \colhead{($\rho_{\earth}$)}
		  }
\startdata
b & $0.66 \pm 0.56$ & ${0.68}^{+0.67}_{-0.58}$ \\
c & $1.17 \pm 0.53$ & ${1.21}^{+0.68}_{-0.57}$ \\
d & $0.89 \pm 0.60$ & ${0.92}^{+0.73}_{-0.63}$ \\
e & $0.80 \pm 0.76$ & ${0.83}^{+0.90}_{-0.79}$ \\
f & $0.60 \pm 0.17$ & ${0.62}^{+0.23}_{-0.19}$ \\
g & $0.94 \pm 0.63$ & ${0.97}^{+0.77}_{-0.66}$ \\
\enddata
\tablenotetext{a}{$\rho_{\earth}=5.51 \text{g cm}^{-3}$}
\tablecomments{No mass and therefore density estimate for TRAPPIST-1h is available from \citet{Gillon2017}.}
\end{deluxetable}

Finally, we calculate density errors for the TRAPPIST-1 planets due to stellar contamination. Integrating the contamination spectra presented in the center panel of Figure~\ref{fig:TRAPPIST1} over the \replaced{I-band}{Spitzer/IRAC 4.5-$\micron$} bandpass, we find an integrated systematic radius error of 
\replaced{$\delta (R_{p}) = 4.0^{+3.8}_{-3.7} \%$}
{$\delta (R_{p}) = 1.1^{+2.5}_{-1.0} \%$},
in which the quoted value is the mean and the error refers to the 68\% dispersion in modeling results. These values translate to density errors of 
\replaced{$\delta(\rho) = -12^{+11}_{-12} \%$}
{$\Delta(\rho) = -3^{+3}_{-8} \%$};
in other words, overestimating the planetary radius leads to an underestimate of the planetary density. Such a systematic error in density measurements could lead to overestimates of the volatile content of the TRAPPIST-1 planets. We provide updated densities for the six innermost TRAPPIST-1 planets in Table~\ref{tab:TRAPPIST1}, adjusting the values reported by \citet{Gillon2017} for the density error due to stellar contamination. Our analysis suggests that stellar contamination may be partially responsible for the relatively low densities reported for planets in this system. More generally, this effect will be an important consideration when selecting targets for characterization follow-up from among a photometrically detected sample.

\section{Conclusions}
\label{sec:conclusions}

We have presented an examination of stellar contamination in \added{visual and near-infrared (0.3--5.5 $\micron$)} transmission spectra of M dwarf exoplanets using model photospheres for M0--M9 dwarf stars with increasing levels of spots and faculae. Our key findings are the following.

\begin{enumerate}

\item
For a given spot covering fraction, larger spots will produce a larger observed variability amplitude than that of smaller spots. Constraining the typical spot size for exoplanet host stars is therefore crucial, since it will mediate the relationship between spot covering fraction and observed variability amplitude. 

\item
The relationship between spot covering fraction and observed variability amplitude is non-linear, scaling generally like a square root relation (Equation~\ref{eq:scaling_relation}) with a coefficient $0.02 < C < 0.11$ that depends on spot contrast and size. As such, previous corrections that have assumed a linear relationship between these variables likely underestimate the true spot covering fraction and, therefore, spectral contamination due to unocculted spots. We will explore this further in an upcoming paper.

\item
In contrast to the low levels of spot covering fractions ($\sim$1\%) used in the literature, we show that a given variability amplitude corresponds to a wide range of spot and faculae covering fractions. For example, assuming spot sizes similar to those of sunspots, we find a typical variability amplitude may correspond to spot covering fractions $0.14 > f_{spot} > 0.44$ and faculae covering fractions $0.56 > f_{fac} > 0.72$ for M0 dwarfs and $0.01 > f_{spot} > 0.24$ and $0.08 > f_{fac} > 0.72$ for M9 dwarfs. These wide ranges correspond to a similarly large uncertainty in the level of stellar contamination present in transmission spectra.

\item
In stars with very large spots, the stellar contamination signal \added{in the 0.3--5.5 $\micron$ wavelength range} can be dominated by the contribution from unocculted faculae, thereby decreasing observed transit depths. However, for stars with solar-like spots and the same observed variability, we find facular coverage to be so widespread that faculae represent the dominate component of the photosphere and are not likely to contribute to a heterogeneity signal.

\item
Depending on spot size, we find the stellar contamination signal can be more than $10 \times$ larger than transit depth changes expected for atmospheric features in rocky exoplanets. Stellar contamination is most problematic for 1) stars with solar-like spots, 2) early M dwarf spectral types, and 3) molecules such as $O_{2}$ with absorption bands at relatively short wavelengths.

\item
We show that the stellar contamination is not limited to visual wavelengths but can also be very significant in the near-infrared bands, likely affecting upcoming James Webb Space Telescope spectroscopy of transiting exoplanets. \added{In the case of a transiting Earth-twin, we find unocculted giant spots can alter the strength of planetary absorption features in transmission spectra by $\sim 20\%$ and unocculted spots and faculae can produce features greater than 30 ppm for spectral types M3V and later.}

\item
We find that stellar spectral contamination across photometric bands due to unocculted starspots with sizes comparable to large sunspot groups can lead to significant errors in radius and therefore density, which may in turn lead to overestimates of planetary volatile content and introduce a stellar-type dependent bias in the density distribution of small planets. This result motivates a possible re-assessment of mass-radius relationships derived from \textit{Kepler} data as well as calls for caution when interpreting upcoming broad-band \textit{TESS} photometry-based exoplanet density measurements.

\item
In the case of the TRAPPIST-1 system, we find spot covering fractions
\replaced{$f_{spot} = 12^{+10}_{-11}\%$}
{$f_{spot} = 8^{+18}_{-7}\%$}
to be consistent with the variability reported by \citet{Gillon2016} when considering variability due to both spots and faculae. The associated stellar contamination signals \added{in the optical and near-infrared} alter transit depths at wavelengths of interest for planetary atmospheric species by roughly 1--15 $\times$ the strength of the planetary feature, significantly complicating James Webb Space Telescope follow-up observations of this system. Similarly, stellar contamination can cause bulk densities of the TRAPPIST-1 planets to be underestimated by 
\replaced{$12^{+11}_{-12} \%$}
{$3^{+8}_{-3} \%$},
leading to overestimates of their volatile contents.
\end{enumerate}

For TRAPPIST-1 and exciting M dwarf exoplanet hosts in general, tighter constraints on spot and faculae covering fractions are crucial for correct interpretations of high-precision \added{visual and near-infrared} transmission spectra from low-mass exoplanets. Stellar contamination is likely to be a limiting factor for detecting biosignatures in transmission spectra of habitable zone planets around M dwarfs. Conversely, exoplanet transit observations in multiple bands can be utilized as a spatial probe to infer the properties of stellar surface heterogeneities. This is particularly critical to achieving an accurate picture of the evolution of exoplanet system properties from young, active systems to those with older, more quiescent stellar hosts. In preparation for precise \textit{JWST} observations probing for molecular features and potential biosignatures such as oxygen, water, and methane from small exoplanets orbiting M dwarfs, we encourage the community to work towards an equally precise understanding of the stellar photospheres providing the light source for transit observations. 

\acknowledgments

B.R. acknowledges support from the National Science Foundation Graduate Research Fellowship Program under Grant No. DGE-1143953. D.A. acknowledges support from the Max Planck Institute for Astronomy, Heidelberg, for a sabbatical visit. \added{This research has made use of NASA's Astrophysics Data System and the Python modules SciPy, NumPy, and Matplotlib.} The results reported herein benefited from collaborations and/or information exchange within NASA's Nexus for Exoplanet System Science (NExSS) research coordination network sponsored by NASA's Science Mission Directorate.  \added{The National Solar Observatory is operated by AURA under a cooperative agreement with the National Science Foundation.}

\bibliography{main}



\end{document}